\documentclass[aip, floatfix, nobibnotes, reprint, superscriptaddress]{revtex4-1}

\usepackage[english]{babel}
\usepackage[utf8]{inputenc}
\usepackage[T1]{fontenc}
\usepackage{amsmath}
\usepackage{amssymb}
\usepackage{amsfonts}
\usepackage{graphicx}
\usepackage{relsize}
\usepackage{bm}
\usepackage{multirow}
\usepackage{bigints}
\usepackage[varg]{txfonts}
\usepackage{bbm}
\usepackage{listings}
\usepackage[caption=false]{subfig}
\usepackage[hmargin=2.5cm, vmargin=4cm]{geometry}
\graphicspath{{figures/}}
\usepackage{url}
\usepackage{longtable}
\usepackage{braket}
\usepackage{textcomp}
\usepackage{float}
\usepackage{bbold} 
\usepackage{comment}   
\usepackage{suffix}
\usepackage[dvipsnames]{xcolor} 
\usepackage[normalem]{ulem}

\newcommand{\papertitle}{Relative energies without electronic perturbations via Alchemical Integral Transform}

\begin{document}

\title{\papertitle}

\author{Simon León Krug}
\affiliation{University of Vienna, Computational Materials Physics, Kolingasse 14-16, 1090 Vienna, Austria}
\affiliation{Machine Learning Group, Technische Universit\"at Berlin and Institute for the Foundations of Learning and Data, 10587 Berlin, Germany}

\author{Guido Falk von Rudorff}
\affiliation{University of Vienna, Computational Materials Physics, Kolingasse 14-16, 1090 Vienna, Austria}
\affiliation{Institute for Pure and Applied Mathematics (IPAM), University of California, Los Angeles, 460 Portola Plaza, Los Angeles, CA 90095, USA}

\author{O. Anatole von Lilienfeld}
\email{anatole.vonlilienfeld@utoronto.ca}
\affiliation{Machine Learning Group, Technische Universit\"at Berlin and Institute for the Foundations of Learning and Data, 10587 Berlin, Germany}
\affiliation{Vector Institute for Artificial Intelligence, Toronto, ON, M5S 1M1, Canada}
\affiliation{Departments of Chemistry, Materials Science and Engineering, and Physics, University of Toronto, St. George Campus, Toronto, ON, Canada}

\date{\today}

\begin{abstract}
We show that the energy of a perturbed system can be fully recovered from the unperturbed system's electron density. 
We derive an alchemical integral transform by parametrizing space in terms of transmutations, the chain rule and integration by parts. Within the radius of convergence, the zeroth order yields the energy expansion at all orders, 
restricting the textbook statement by Wigner that the $p$-th order wave function derivative is necessary to describe the $(2p+1)$-th energy derivative. Without the need for derivatives of the electron density, this allows to cover entire chemical neighborhoods from just one quantum calculation instead of single systems one-by-one. 
Numerical evidence presented indicates that predictive accuracy is achieved in the range of mHa for the harmonic oscillator or the Morse potential, and in the range of machine accuracy for hydrogen-like atoms.
Considering iso-electronic nuclear charge variations by one proton 
in all multi-electron atoms from He to Ne,
 alchemical integral transform based estimates of the relative energy
 deviate by only few mHa from corresponding Hartree-Fock reference numbers. 
\end{abstract}

\maketitle

\section{Introduction}


The energy of a system is central to quantum mechanics, and can be obtained as the solution to the eigenproblem of the Hamiltonian.
Solving the electronic Schrödinger equation for real compounds arguably constitutes the most severe bottleneck for our understanding of trends among chemical systems. 
Unfortunately, the number of conceivably stable materials and molecules is colossal, making brute-force enumeration attempts prohibitive. 
One possible alternative is to connect the solutions of two chemically distinct systems via continuous interpolation of nuclear charges, aka~"alchemical" changes \cite{PRL_compalchemy,lilienfeld_variational}, as introduced by E.~B.~Wilson already in 1962~\cite{bright_wilson}. 
Alchemical Perturbation Density Functional Theory~(APDFT)~\cite{von_Rudorff_2020} 
couples the Hamiltonians of two iso-electronic systems with a 
mixing parameter $\lambda$ and generates approximations of relative energies as a perturbative series \cite{weigend_firstalchemy, sheppard_firstTaylorexpansion}.
For sufficiently small changes in chemical composition, 
this series has recently been demonstrated to converge beyond numerical precision~\cite{rudorff2021arbitrarily}.
By virtue of the Hellmann-Feynman theorem (cf.~Eq.~\ref{eq:HellmannFeynmanApplied}), 
this perturbative expansion relies exclusively on electron density and derivatives w.r.t.~$\lambda$~\cite{von_Rudorff_2020}. 
Perturbations of the electron density at any order are typically calculated explicitly (cf.~Eq.~\ref{eq:original_expansion}) via e.g. finite differences~\cite{johnkeith,von_Rudorff_2020} or coupled-perturbed equations~\cite{balawender_alchemDer, coupled_pert_wolinski, dunlap_CPHF,giorgio}, imposing significant computational cost. 
Recent implementations of automatic differentiation~\cite{ad_margossian} in numerical libraries~\cite{tensorflow, pytorch, keras, theano}, and even in dedicated quantum chemistry software~\cite{diffiqult, pennylane, dqc_pennylane,dqc_kasim}, hold substantial
promise to accelerate APDFT based exploration campaigns of materials compounds space \cite{ceder1998predicting}.



In computational catalysis, alchemy has been used for binding energy predictions with great effect up to first perturbative order\cite{Alchemy_bindingenergies}. A density-derivative-free approach would achieve additional accuracy by naturally including higher orders without an increase in compuational cost.

Here, the initial, unperturbed electron density is shown to include sufficient quantum mechanical information to recover the relative energy with respect to any iso-electronic final system -- as long as the alchemical expansion converges.
The calculation of density-derivatives therefore becomes unnecessary. 
This represents a substantial improvement over Wigner's textbook statement that the $p$-th order wave function derivative describes the $(2p+1)$-th energy derivative \cite{2p+1}. 
Below, we provide the derivation and its discussion, followed by numerical results. 
In particular, we have considered iso-electronic relative energies among hydrogen-like atoms, and among multi-electron atoms covering nuclear charges from~H to~Ne.


\section{Theory}

The starting point of our formulation is APDFT~\cite{von_Rudorff_2020}. Consider any two iso-electronic systems with their electronic Hamiltonians $\hat{H}_A$ and $\hat{H}_B$ with corresponding external potentials $v_A$ and $v_B$ which are connected via a linear transformation such that $\hat{H}(\lambda) = \hat{H}_A (1-\lambda)+ \hat{H}_B \lambda$ and $v(\lambda) = v_A (1-\lambda)+ v_B \lambda$.
Given a general electron density $\rho(\lambda, \bm{r})$, which yields the electron density at any $\lambda \in [0,1]$, 
the first order derivative according to the Hellmann-Feynman theorem corresponds to 
\begin{align}
    \label{eq:HellmannFeynmanApplied}
	\frac{\partial E(\lambda)}{\partial \lambda} &= \bra{\Psi_{\lambda}}\hat{H}_B - \hat{H}_A\ket{\Psi_{\lambda}}
	=\int_{\mathbb{R}^3} d\bm{r}\,\, \Delta v \,\, \rho(\lambda, \bm{r})
\end{align}

\noindent
with difference in external potentials $\Delta v = v_B - v_A$. 

We express $E_B=\bra{\Psi_{B}}\hat{H}_B\ket{\Psi_{B}}$ by perturbatively expanding $E(\lambda) = \bra{\Psi_{\lambda}}\hat{H} (\lambda) \ket{\Psi_{\lambda}}$ at $\lambda = 0$. 
Inserting Eq.~\ref{eq:HellmannFeynmanApplied} into the perturbative expansion
\begin{align}
	\Delta E := E_B - E_A &= \sum_{p = 1}^{\infty} \frac{1}{p!} \frac{\partial E(\lambda)}{\partial \lambda} \bigg\vert_{\lambda = 0} \, \Delta \lambda ^p \label{eq:energy_taylor} \\
	&= \sum_{p = 1}^{\infty} \frac{1}{p!} \int_{\mathbb{R}^3} d\bm{r}\, \Delta v \, \frac{\partial^{p-1} \rho(\lambda, \bm{r})}{\partial \lambda^{p-1}} \bigg\vert_{\lambda = 0}
	\label{eq:original_expansion}
\end{align}

\noindent
with energy difference~$\Delta E = E_B - E_A$. This formula can be rewritten by transferring the $\lambda$-dependency of the general $\rho(\lambda, \bm{r})$ to a parametrization of the spatial coordinates~$\bm{r}(\lambda)$. 

As a simplified introduction to the concept and its subsequent generalization, let us first consider the one-dimensional case (in $\mathfrak{r}$) of Eq. \ref{eq:original_expansion} for the first non-linear order ($p=2$) with $\mathfrak{r}_A := \mathfrak{r}(\lambda = 0)$.
Parametrizing $\mathfrak{r} \rightarrow \mathfrak{r}(\lambda)$ with Jacobian $J$:
\begin{align}
    \Delta E^{(2)} &= \frac{1}{2} \int_{\mathbb{R}} d\mathfrak{r}(\lambda) \,J\,  \Delta v (\mathfrak{r}(\lambda)) \,\frac{\partial \rho(\lambda, \mathfrak{r}(\lambda))}{\partial \lambda} \bigg\vert_{\lambda = 0}
\end{align}
Rewriting $\rho(\lambda, \mathfrak{r}(\lambda)) = \rho(\mathfrak{r}^{-1}(\mathfrak{r}(\lambda)), \mathfrak{r}(\lambda)) =: \tilde{\rho}(\mathfrak{r}(\lambda))$ and using the chain rule:
\begin{align}
    \Delta E^{(2)} &= \frac{1}{2} \int_{\mathbb{R}} d\mathfrak{r}(\lambda) \,J\,  \Delta v (\mathfrak{r}(\lambda)) \,\frac{\partial \tilde{\rho}(\mathfrak{r}(\lambda))}{\partial \mathfrak{r}(\lambda)} \frac{\partial \mathfrak{r}(\lambda)}{\partial \lambda} \bigg\vert_{\lambda = 0}
\end{align}

When integrating by parts, all limit terms equal zero since the electron density vanishes at infinite distance and $\partial_{\mathfrak{r}}J = 0$. Inserting $\lambda = 0$ wherever possible ($J\vert_{\lambda=0} = 1$):
\begin{align}
    \Delta E^{(2)} &= -\frac{1}{2} \int_{\mathbb{R}} d\mathfrak{r}_A \, \rho_A(\mathfrak{r}_A)\, \frac{\partial \Delta v(\mathfrak{r}_A)}{\partial \mathfrak{r}_A} \,\frac{\partial \mathfrak{r}(\lambda)}{\partial \lambda} \bigg\vert_{\lambda = 0}
\end{align}
\noindent

\textit{Vide infra} for our general \textit{Ansatz} of~$\partial_{\lambda} \mathfrak{r}\vert_{\lambda=0}$ for all orders~(Eq.~\ref{eq:r_lambda_full}) and of the full parametrization (Eq.~\ref{eq:theta_wi_molecules}).

We now generalize this by applying Fa\`a di Bruno's formula (repeated chain rule) for composite functions with a vector argument~\cite{Mishkov2000} to obtain any derivatives of $\rho$~w.r.t.~$\bm{r}(\lambda)$ and $\bm{r}(\lambda)$~w.r.t.~$\lambda$. Again, all derivatives of $\rho$~w.r.t.~$\bm{r}(\lambda)$ equal zero through repeated partial integration, as all spatial derivatives of the electron density vanish at infinite distance (cf.~Supplemental Material for the detailed derivation):

\begin{widetext}

\begin{align}
    \label{eq:Delta_E_coeff_final}
    \Delta E=& \int_{\mathbb{R}^3} d\bm{r}_A\, \rho_A \left( \bm{r}_A \right) \, \, \mathcal{K} \left( \bm{r}_A, v_A, v_B \right)\\
    \mathcal{K} \left( \bm{r}_A, v_A, v_B \right) = &\sum_{p = 1}^{\infty} \frac{1}{p} \sum_{S_p} \frac{\partial^{\mu_x + \mu_y + \mu_z} \Delta v (\bm{r}_A)}{\partial x_A^{\mu_x} \partial y_A^{\mu_y} \partial z_A^{\mu_z}}
    \left[\prod_{i = 1}^{p-1} \frac{ \left( \sum_{w \in \lbrace x,y,z \rbrace} \theta_{w,i} \right)^{k_i}}{k_i!} \right] \label{eq:Delta_E_coeff_final_K_withTheta} \\
    \theta_{w,i} := & - \frac{1}{i!} \frac{\partial^i w (\lambda)}{\partial \lambda^i} \bigg\vert_{\lambda = 0}\\
    \label{eq:Sp}
    S_p :=& \left\lbrace \mu_x, \mu_y, \mu_z, k_1, \dots, k_{p-1} \in \mathbb{N}_0 \, \Bigg\vert \, p-1 = \sum_{i=1}^{p-1} i \cdot k_i , \, \mu_x + \mu_y + \mu_z = \sum_{i=1}^{p-1} k_i\right\rbrace
\end{align}
\end{widetext}

\begin{figure*}[t]
    \centering
    \includegraphics[width=0.97\textwidth]{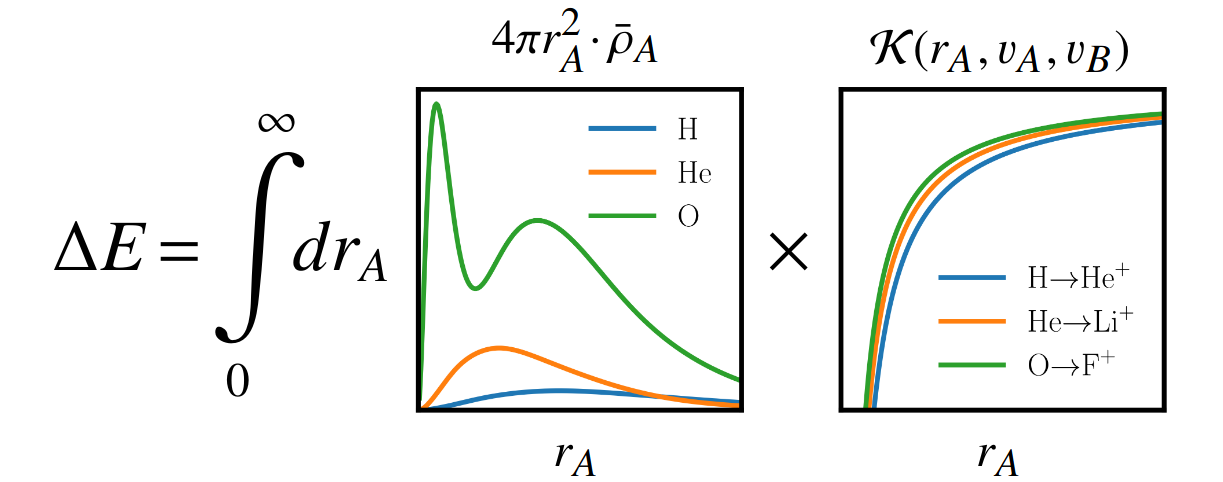}
    \caption{Visualization of the Alchemical Integral Transform (AIT) for radial electron densities and kernel $\mathcal{K}$ corresponding to iso-electronic alchemical changes, exemplified for atoms H $\rightarrow$ He$^+$, He $\rightarrow$ Li$^+$, O $\rightarrow$ F$^+$.}
    \label{fig:visualized_eq}
\end{figure*}

Here, we denote the three spatial variables $w_A \in \lbrace x_A,y_A,z_A \rbrace$. 
The subscript $A$ is needed to discriminate between $w_A$ and the parametrization of the general coordinates $w(\lambda)$, including its derivatives in $\theta_{w,i}$ which emerge from repeated use of the chain rule. 
Note that $\mathcal{K}$ now includes derivatives of general spatial coordinates $w(\lambda)$ w.r.t.~$\lambda$ which is akin to morphing space, or rather repositioning the values of~$\rho_A$. 
Although both initial and final system are described by stationary Hamiltonians, this parametrization in $\lambda$ suggests a mathematical treatment in four dimensions~$\lambda,x,y,z$ rather than three, where (possibly non-integer) nuclear charges~$Z$ enjoy attention at par with the coordinates, reminiscent of the aforementioned four-dimensional electron density introduced by Wilson~\cite{bright_wilson}.
Note that the energy difference $\Delta E$ in Eq.~\ref{eq:Delta_E_coeff_final} does not depend on a general~$\rho(\lambda, \bm{r})$ or its derivatives anymore but only on the initial~$\rho_A$ and a function~$\mathcal{K}(\bm{r},v_A, v_B)$. 
Furthermore, $\Delta E$ is not perturbative in $\rho(\lambda, \bm{r})$ anymore, but rather in $\Delta v$ which is known analytically. 
Even without an explicit expression for $\theta_{w,i}$, this heavily constraints the statement of Wigner's $(2p+1)$-theorem of perturbation theory: "the first $p$ perturbative eigenfunctions are sufficient to produce the first $2p+1$ eigenenergy derivatives."~\cite{2p+1} Now, the zeroth order ($p = 0$) generates \textit{all} energy derivatives and hence, renders density derivatives obsolete -- within the radius of convergence.

Emphasizing its alchemical characteristics and applicability for calculation of energy differences without density derivatives, 
we henceforth dub this method Alchemical Integral Transform~(AIT) since Eq.~\ref{eq:Delta_E_coeff_final} constitutes an integral transform with non-linear kernel $\mathcal{K}(\bm{r}_A, v_A(\bm{r}_A), v_B(\bm{r}_A))$. 
Assuming the parametrization $\bm{r} (\lambda)$ to be analytical on $[0,1]$ and bijective in an environment of $\lambda = 0$ (cf.~Eq.~\ref{eq:r_lambda_full}), AIT is rigorous for iso-electronic changes. 
The AIT in Eq.~\ref{eq:Delta_E_coeff_final} proves the hypothesis that the unperturbed initial density might suffice for calculating relative energies, as introduced and empirically corroborated in 2009 using non-linear interpolations in $\lambda$~\cite{accurate_abinitio}. 
Non-linear $\lambda$-interpolations were also already used within the context of free energy perturbation estimates relying on molecular dynamics based sampling~\cite{HigherOrderAlchemicalDerivatives_SmithGunsteren1994}.
Relative energies according to AIT are illustrated in Fig. \ref{fig:visualized_eq} for the alchemical increase in nuclear charge of atoms.

We now propose the following \textit{Ansatz} for $\theta_{w,i}$ based on dimensional arguments,
\begin{align}
    \label{eq:theta_wi_molecules}
    \theta_{w,i} =& \, \left( 1 - \frac{v_B(\bm{r}_A)}{v_A(\bm{r}_A)} \right)^i \cdot w_A  \quad \text{,}
\end{align}
where $v_A$ and $v_B$ correspond to the known external potentials of \textit{any} pair of iso-electronic systems.
As we gathered all derivatives of the parametrization $\bm{r}(\lambda)$ at $\lambda = 0$ through Eq.~\ref{eq:theta_wi_molecules}, we can construct $\bm{r}(\lambda)$ via its Maclaurin series:
\begin{align}
    \label{eq:r_lambda_maclaurin}
    \bm{r}(\lambda) =& \sum_{i = 0}^{\infty} \frac{\partial^i \bm{r} (\lambda)}{\partial \lambda^i} \bigg\vert_{\lambda = 0}\frac{\lambda^i}{i!} = \bm{r}_A - \sum_{i = 1}^{\infty} \theta_{i,\bm{r}} \, \lambda^i\\
    \label{eq:r_lambda_full}
    =& \, \left(2 - \frac{v_A}{(v_B - v_A)\lambda - v_A} \right) \bm{r}_A
\end{align}
This parametrization fulfills both bijectivity and invertibility in $\lambda$ on $[0,1]$ at almost every $\bm{r}_A$. 
By explicitly stating $\theta_{w,i}$, Eqs.~\ref{eq:Delta_E_coeff_final} to~\ref{eq:Sp} provide not just an existence proof for a restriction of Wigner's $(2p+1)$-theorem, but now serve as a method capable of quantitative predictions.
Matching relative energies of analytically solvable models like the hydrogen-like atom, the quantum harmonic oscillator, the Morse potential or the Dirac well using AIT (cf. Supplemental Material) provides strong evidence that Eqs.~\ref{eq:theta_wi_molecules} to~\ref{eq:r_lambda_full} hold for general systems.

From Eqs.~\ref{eq:Delta_E_coeff_final_K_withTheta}, \ref{eq:theta_wi_molecules} and \ref{eq:r_lambda_maclaurin}, a convergence condition can be read off:
\begin{align}
    \label{eq:naive_convergence}
    \bigg\vert 1 - \frac{v_B(\bm{r}_A)}{v_A(\bm{r}_A)} \bigg\vert < 1 \qquad \forall \bm{r}_A \in {\mathbb{R}^3}
\end{align}
The \textit{general} convergence behavior of the alchemical Taylor expansion is not rigorously proven except for special cases~\cite{von_Rudorff_2020}, but numerical evidence points towards large convergence radii of the alchemical Taylor expansion even for large changes in external potential~\cite{rudorff2021arbitrarily}, also crucially impacted by the quality of the basis set (\textit{vide infra}). 
For $v_B$ exceeding $v_A$ in mono-atomic systems beyond this convergence criterion, on can observe the resulting divergence in the Supplemental Material (Fig.~\ref{fig:monoatomics_proof}).

AIT and the original perturbative energy expansion match order-wise. This enables the evaluation of energy-perturbations $\partial^p_{\lambda} E(\lambda) \vert_{\lambda=0}$ at selected orders $p$ and gives access to all proportional quantities, e.g. the alchemical potential~$\partial_{Z_A} E_A$ and the spatial gradient~$\partial_{\bm{R}_A} E_A$ at first order ($p=1$), or the alchemical hardness~$\partial^2_{Z_A}E_A$ (\textit{vide infra}) and the spatial Hessian~$\partial^2_{\bm{R}_A} E_A$ at second order ($p=2$)~\cite{lilienfeld_tuckerman}. Analytical higher order energy derivatives have already been presented in the context of Kohn-Sham-DFT~\cite{balawender_alchemDer}.

We want to point out that AIT is not limited to Coulombic potentials in three dimensions, but can be shown (cf. Supplemental Material) to be applicable for the Dirac delta well, the quantum harmonic oscillator and the Morse potential, each in one dimension. At the same place, we show that AIT can be generalized to periodic systems in arbitrary dimensions. This provides a strong indication that AIT holds for a wide range of general initial and final potentials, which are not necessarily of the same functional form, but can be of arbitrary finite dimensionality. 

\section{Results}



\begin{figure}[t!]
    \centering
    \includegraphics[width=0.49\textwidth]{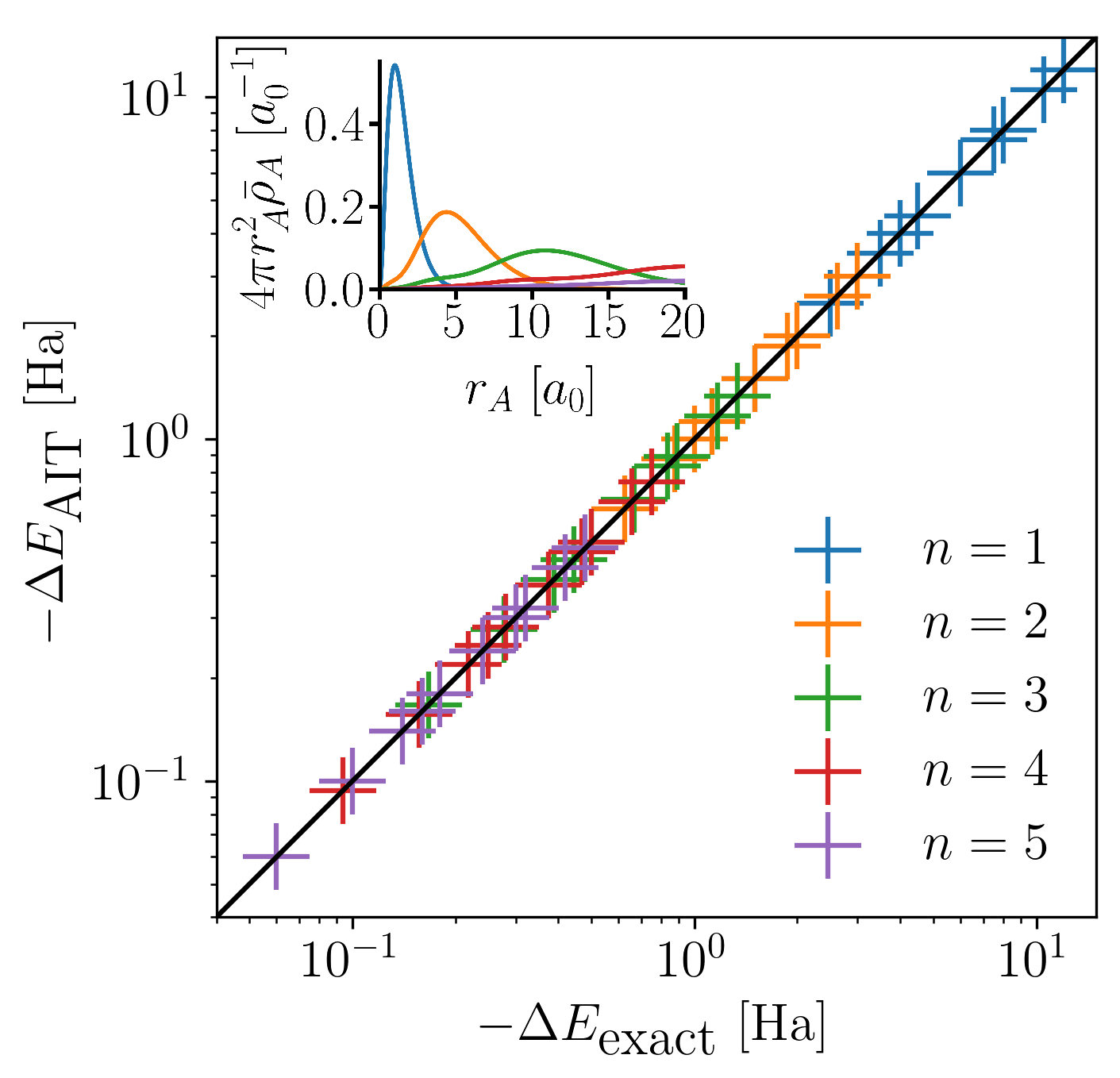}
    \caption{AIT for the hydrogen-like atom: the analytically known energy $\Delta E_{\text{exact}}$ scattered against the AIT energy $\Delta E_{\text{AIT}}$, up to and including fifth perturbation order $p$ for different quantum numbers $n$ where initial and final nuclei $Z_A,Z_B \in \lbrace 1,2,3,4,5 \rbrace$, (H, He$^+$, Li$^{2+}$, Be$^{3+}$, B$^{4+}$), and $Z_A < Z_B$. The numbers agree within $10^{-14}$ Ha. Radial electron densities of H for different $n$ are shown in the inset.}
    \label{fig:hydrogen_proof}
\end{figure}

\begin{figure}[t]
    \centering
    \includegraphics[width=0.48\textwidth]{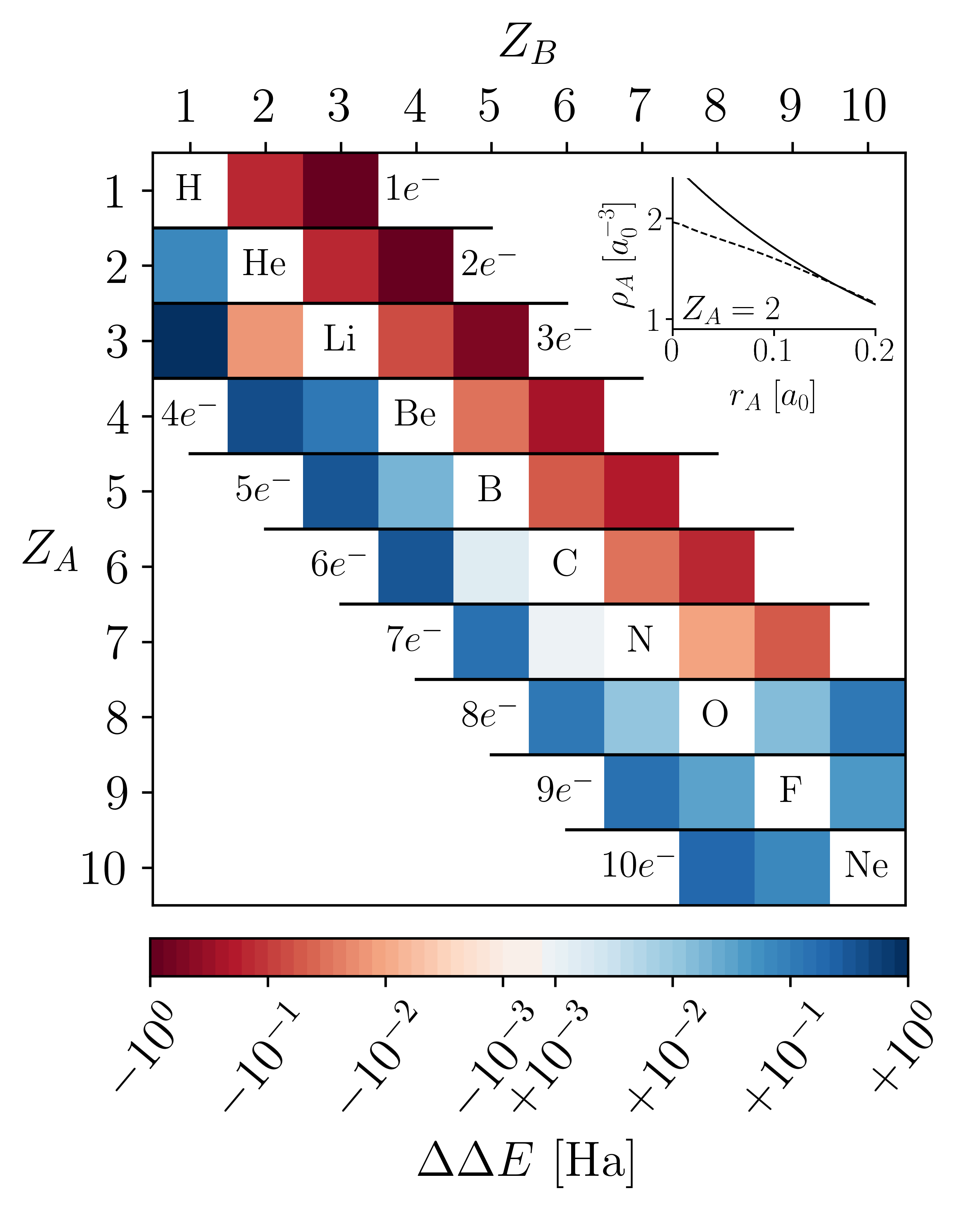}
    \caption{AIT based predictions for multi-electron atoms: Error $\Delta \Delta E = \Delta E_{\text{SCF}} - \Delta E_{\text{AIT}}$ between unrestricted Hartree-Fock SCF energy difference $\Delta E_{\text{SCF}}$ and AIT estimate $\Delta E_{\text{AIT}}$ up to and including fifth perturbation order $p$. Initial electron densities were obtained for $Z_A \in \lbrace 1, \dots,10\rbrace$ (H to Ne), and final nuclear charges considered include $Z_B = Z_A \pm dZ$ with $dZ \in \lbrace 1,2 \rbrace$. The inset shows electron densities of the hydrogen-like atom at $Z_A = 2$ (He$^+$) from the analytical solution (solid) and SCF-calculations (dashed).}
    \label{fig:deviation_monoatomic}
\end{figure}

In the following, and without any loss of generality, we restrict ourselves to the case of atoms, $v_B/v_A \rightarrow Z_B/Z_A$,
which we show to be correct up to numerical precision in case of the hydrogen-like atom (Fig.~\ref{fig:hydrogen_proof}) and to be applicable to multi-electron atoms (Fig.~\ref{fig:deviation_monoatomic}). 
In the hydrogen-like atom, energy, wave function, and electron density are analytically known. 
From an initial atom~$Z_A$, we transmute to some final~$Z_B$ (in atomic units) with principal quantum number $n$:
\begin{align}
    \label{eq:hydrogenlike-exact_energies}
    \Delta E_{\text{exact}} = -\frac{Z_B^2 - Z_A^2}{2n^2}
\end{align}
AIT 
simplifies drastically in radially symmetric systems:
\begin{align}
	\Delta E_{\text{AIT}}= & \sum_{p = 1}^{\infty} \left( \frac{1}{p}\sum_{T_p} \frac{(-1)^{\mu_r} \cdot \mu_r!}{\prod_i^{p-1} k_i!} \right)  (-Z_B + Z_A) \left( 1 - \frac{Z_B}{Z_A} \right)^{p-1} \notag \\
	&\times \int\limits_0^{\infty} dr_A \, 4\pi r_A \, \bar{\rho}_A(r_A, n, Z_A) \label{eq:Delta_E_coeff_radial}
\end{align}
\begin{align}
    T_p :=& \left\lbrace \mu_r, k_1, \dots, k_{p-1} \! \in \! \mathbb{N}_0 \, \Bigg\vert p\!-\!1 \!= \! \sum_{i=1}^{p-1} i \cdot k_i , \, \mu_r \! = \!\sum_{i=1}^{p-1} k_i\right\rbrace
\end{align}
The derivation and the spherically averaged electron density $\bar{\rho}_A$ are given in the Supplemental Material. 
In Fig. \ref{fig:hydrogen_proof}, 
$\Delta E_{\text{exact}}$ is compared with $\Delta E_{\text{AIT}}$ up to and including fifth perturbation order $p$, 
fifth state $n$, and for all transmutations between the first five elements in the periodic table (H, He$^+$, Li$^{2+}$, Be$^{3+}$, B$^{4+}$) with $Z_A < Z_B$. 
We find the two results to agree with numerical precision (10$^{-14}$~Ha).

As both energy expressions (Eqs.~\ref{eq:hydrogenlike-exact_energies} and~\ref{eq:Delta_E_coeff_radial}) match, we find connections between nuclear charges and the radial expectation value of the electron density:
\begin{align}
    \frac{Z_B + Z_A}{2n^2} =& \sum_{p = 1}^{\infty} \left( \frac{1}{p}\sum_{T_p} \frac{(-1)^{\mu_r} \cdot \mu_r!}{\prod_i^{p-1} k_i!} \right) \cdot \left( 1 - \frac{Z_B}{Z_A} \right)^{p-1} \notag \\
    & \times \int\limits_0^{\infty} dr_A \, 4\pi r_A \, \bar{\rho}_A(r_A, n, Z_A)
\end{align}
and in the limit of $Z_B \rightarrow Z_A$,
\begin{align}
    \label{eq:gausslaw}
    \frac{Z_A}{4\pi n^2} =& \int\limits_0^{\infty} dr_A \, r_A \cdot \bar{\rho}_A(r_A, n, Z_A) =: \left< \, \bar{\rho}_A \right>_{r_A}
\end{align}

\noindent
with radial expectation value $\left< \dots \right>_{r_A}$. This can also be derived from the analytic electron density expression (cf.~Supplemental Material) which confirms the equivalence in this particular case. 
Furthermore, the alchemical hardness~$\eta_{\text{al}}$ can be obtained via the second perturbative order ($p=2$) of Eq.~\ref{eq:Delta_E_coeff_radial}:
\begin{align}
    \Delta E^{(2)}_{\text{AIT}} &=\frac{1}{2} \frac{\partial^2 E_A}{\partial Z_A^2} \, (Z_B - Z_A)^2  \\
    &=\frac{Z_B - Z_A}{2} \left( 1 - \frac{Z_B}{Z_A} \right) \int\limits_0^{\infty} dr_A \, 4\pi r_A \, \bar{\rho}_A(r_A, n, Z_A)\\
    \Rightarrow \eta_{\text{al}} &=  \frac{\partial^2 E_A}{\partial Z_A^2} = -\frac{4\pi}{Z_A} \left< \, \bar{\rho}_A \right>_{r_A}
\end{align}
This relation holds even for multi-electron atoms.

AIT for systems with more than one electron is inherently more challenging because $\Delta E_{\text{AIT}}$ depends on $\rho_A$ and numerical electron densities of a multi-electron system are only known approximately (cf. Fig.~\ref{fig:monoatomics_proof} in the Supplemental Material). 
Thus, accuracy suffers from approximations made to the electron correlation problem, as well as from incomplete basis set effects (Pulay-forces~\cite{pulay}), as shown in Refs.~\onlinecite{von_Rudorff_2020,giorgio_basisset}. 
Approaching the complete basis set limit with larger basis sets like \texttt{def2-TZVP} or \texttt{cc-pV5Z} reduces the latter error .

Our comparison of $\Delta E_{\text{AIT}}$ and $\Delta E_{\text{SCF}}$ for different basis sets with the basis functions fixed to the individual initial atom indicates that accuracy benefits the most from using Hartree-Fock with the \texttt{def2-TZVP} basis functions of Xe (details given in the Supplemental Material).
To make this selection, we have tested different combinations of basis sets 
(of the families Pople, Dunning, Dunning Douglas-Kroll, Dunning JK-fitting, Ahlrichs, Lehtola, ANO and STO\cite{bse}), basis functions, and different levels of theory 
in terms of treating electron correlation (HF, ROHF, CISD, CCSD, and FCI).
We always encounter a mean average error (MAE) of at least 50~mHa when performing benchmarks on the elements H to Ne and $\Delta Z = \pm 1$.

Similar discrepancies in accuracy among basis sets for elements have also been seen before for alchemical derivatives involving noble gases~\cite{chemical_space_balawender}. 


For an overarching picture of AIT's accuracy for multi-electron-atom predictions, we have used the best-performing basis set, \texttt{def2-TZVP} with the basis functions of Xe, to quantify the absolute error $\Delta \Delta E = \Delta E_{\text{SCF}} - \Delta E_{\text{AIT}}$ for iso-electronic interpolations where $1 \leq Z_A, Z_B \leq 10$. 
Fig. \ref{fig:deviation_monoatomic} displays prediction errors as a heat map for all atoms $Z_A$ from H up to Ne, with
$Z_B = Z_A \pm dZ$ with $dZ \in \lbrace1,2 \rbrace$. 
The number of electrons $N_e$ of the initial atom $Z_A$ always equals its nuclear charge such that the overall charge $Q_A = 0$. A version with non-integer $Z_A, Z_B$ can be found in the Supplemental Material.
The energies of e.g. N$^{-}$ and F$^{+}$ can be estimated based only on the unperturbed electron density of oxygen within $\sim$19 and $\sim$24\,mHa, respectively.
Especially close to the nucleus, $\Delta E_{\text{AIT}}$ becomes very sensitive to errors in electron density as $| \mathcal{K} |$ becomes large (cf. Fig.~\ref{fig:visualized_eq}). 
This sensitivity overtly shows when calculating energy differences between H and another hydrogen-like atom:
while the approximate self-consistent field (SCF)-solution diverges for $Z_B > 2$  
as the basis set has no sufficiently tight basis functions, the analytical solution holds even for $Z_B = 5$ (Fig.~\ref{fig:hydrogen_proof}). This difference in electron densities can be seen in the inset of Fig. \ref{fig:deviation_monoatomic} for He$^+$. Clearly, the SCF solution suffers in accuracy when stacking positive charges in the nucleus because the basis sets were not designed to properly account for such  deformation, as also
recently discussed in Ref.~\onlinecite{jensen_basisset}. 
Furthermore, there appears to be a sudden change in the sign of $\Delta \Delta E$ between~N and~O. This trend is identical for arbitrary perturbation orders $p > 2$, hence we reckon this to be not due to truncation or divergence of the series in Eq.~\ref{eq:Delta_E_coeff_final_K_withTheta}, but instead the quality of electron densities from Hartree-Fock methods.
Increasing the initial density's quality will be crucial to access the chemical neighborhood with better accuracy.

In summary, we have introduced AIT, the Alchemical Integral Transform that turns a converging perturbative expansion of relative energies in electron density derivatives into an analytical expansion in external potentials and space derivatives.
Our \textit{Ansatz} naturally leads to 
accurate energy predictions of atoms with neighboring nuclear charges, as demonstrated for hydrogen-like atoms as well as for all atoms up to Ne.
As a consequence, we could show that only the initial unperturbed electron density is required for AIT, rendering negligible the computational cost for accurate estimates of relative energies of distinct iso-electronic systems. 
In conclusion, all the relevant quantum mechanical information for any iso-electronic system of the same quantum state is already contained in the initial electron density, restricting Wigner's ($2p+1$)-theorem of perturbation theory. 
Consequently, this constitutes a strong hint to apply alchemical methods in general, and AIT in specific, to obtain also relative electron densities.


Future work will deal with other systems, as a generalization to molecules and materials appears to have
great merit to explore vast regions of materials compound space more efficiently based only on few unperturbed electron densities. 
Our particular \textit{Ansatz}, the role of electron correlation treatment, as well as basis sets effects (cf.~magnitude and sign of $\Delta  \Delta E$ in Fig.~\ref{fig:deviation_monoatomic}) and the overall quality of the electron density, might 
also be worthy of further attention.
Especially the latter seems promising to gain accuracy, as an electron density of higher quality may be computationally expensive, but once obtained, enables access to a multitude of (accurate) relative energies.

\section*{Supplementary Material}
See the supplementary material for the derivation of Eqs.~\ref{eq:Delta_E_coeff_final} and~\ref{eq:Delta_E_coeff_final_K_withTheta}, the solutions to the linear Diophantine equations (cf.~Eq.~\ref{eq:Sp}), the functional form of the electron density of the hydrogen-like atom, the derivation of Eq.~\ref{eq:Delta_E_coeff_radial}, an alternative proof of Eq.~\ref{eq:gausslaw}, the application of AIT to toy models (Dirac well, quantum harmonic oscillator, Morse potential, periodic potentials) and details regarding the performance of basis sets in multi-electron atoms.

\section*{Acknowledgements}
We acknowledge discussions with M.~Meuwly, D.~Lemm and H.~Schäfer, as well as support from the European Research Council (ERC-CoG Grant QML). This project has received funding from the European Union's Horizon 2020 research and innovation programme under Grant Agreement \#772834. 

\section*{Conflict of Interest}
The authors have no conflicts to disclose.

\section*{Author Contributions}
\textbf{Simon León Krug}: conceptualization (lead), data curation, formal analysis (lead), investigation (lead), methodology (lead), software, visualization (equal), writing - original draft (lead), writing - review \& editing (supporting).
\textbf{Guido Falk von~Rudorff}: conceptualization (supporting), formal analysis (supporting), investigation (supporting), methodology (supporting), supervision (supporting), visualization (equal), writing - original draft (equal).
\textbf{O.~Anatole von~Lilienfeld}: conceptualization (supporting), formal analysis (supporting), investigation (supporting), methodology (supporting), funding acquisition, project administration, resources, supervision (lead), visualization (equal), writing - review \& editing (lead).

All authors read and approved the final manuscript.

\section*{Data and code availability}
The data that support the findings of this study, namely the data basis of Figs.~\ref{fig:deviation_monoatomic} and~\ref{fig:deviation_monoatomic_nonints}, is openly available on Zenodo and can be found under \url{https://doi.org/10.5281/zenodo.6779769}. The code that produces the findings of this study, in specific the comparisons of the hydrogen-like atom, multi-electron atom, quantum harmonic oscillator and Morse potential, are openly available on GitHub under \url{https://github.com/SimonLeonKrug/pyalchemy}.

\bibliography{refs.bib}{}

\begin{thebibliography}{10}

\bibitem{PRL_compalchemy}
Nicola Marzari, Stefano de~Gironcoli, and Stefano Baroni.
\newblock Structure and phase stability of {$\text{Ga}_x \text{In}_{1-x}
  \text{P}$} solid solutions from computational alchemy.
\newblock {\em Physical review letters}, 72(25):4001--4004, 1994.

\bibitem{lilienfeld_variational}
O.~Anatole von Lilienfeld, Roberto~D. Lins, and Ursula Rothlisberger.
\newblock Variational particle number approach for rational compound design.
\newblock {\em Phys. Rev. Lett.}, 95:153002, Oct 2005.

\bibitem{bright_wilson}
E.~Bright Wilson.
\newblock Four‐dimensional electron density function.
\newblock {\em The Journal of Chemical Physics}, 36(8):2232--2233, 1962.

\bibitem{von_Rudorff_2020}
Guido~Falk von Rudorff and O.~Anatole von Lilienfeld.
\newblock Alchemical perturbation density functional theory.
\newblock {\em Physical Review Research}, 2(2), 5 2020.

\bibitem{weigend_firstalchemy}
Florian Weigend, Claudia Schrodt, and Reinhart Ahlrichs.
\newblock Atom distributions in binary atom clusters: a perturbational approach
  and its validation in a case study.
\newblock {\em The Journal of chemical physics}, 121(21):10380--10384, 2004.

\bibitem{sheppard_firstTaylorexpansion}
Daniel Sheppard, Graeme Henkelman, and O~Anatole von Lilienfeld.
\newblock Alchemical derivatives of reaction energetics.
\newblock {\em The Journal of chemical physics}, 133(8):084104--084104--7,
  2010.

\bibitem{rudorff2021arbitrarily}
Guido~Falk von Rudorff.
\newblock Arbitrarily precise quantum alchemy.
\newblock 2021.

\bibitem{johnkeith}
Emily~A Eikey, Alex~M Maldonado, Charles~D Griego, Guido~Falk von Rudorff, and
  John~A Keith.
\newblock Evaluating quantum alchemy of atoms with thermodynamic cycles: Beyond
  ground electronic states.
\newblock {\em The Journal of chemical physics}, 156(6):064106--064106, 2022.

\bibitem{balawender_alchemDer}
Michał Lesiuk, Robert Balawender, and Janusz Zachara.
\newblock Higher order alchemical derivatives from coupled perturbed
  self-consistent field theory.
\newblock {\em The Journal of Chemical Physics}, 136(3):034104, 2012.

\bibitem{coupled_pert_wolinski}
Krzysztof Wolinski, James~F Hinton, and Peter Pulay.
\newblock Efficient implementation of the gauge-independent atomic orbital
  method for nmr chemical shift calculations.
\newblock {\em Journal of the American Chemical Society}, 112(23):8251--8260,
  1990.

\bibitem{dunlap_CPHF}
B~I Dunlap and J~Andzelm.
\newblock 2nd derivatives of the local-density-functional total energy when the
  local potential is fitted.
\newblock {\em Physical review. A, Atomic, molecular, and optical physics},
  45(1):81--87, 1992.

\bibitem{giorgio}
Giorgio Domenichini and O.~Anatole von Lilienfeld.
\newblock Alchemical geometry relaxation.
\newblock {\em arXiv:2201.07129}, 2022.

\bibitem{ad_margossian}
Charles~C. Margossian.
\newblock A review of automatic differentiation and its efficient
  implementation.
\newblock {\em {WIREs} Data Mining and Knowledge Discovery}, 9(4), Mar 2019.

\bibitem{tensorflow}
Mart\'{i}n Abadi, Ashish Agarwal, Paul Barham, Eugene Brevdo, Zhifeng Chen,
  Craig Citro, Greg~S. Corrado, Andy Davis, Jeffrey Dean, Matthieu Devin,
  Sanjay Ghemawat, Ian Goodfellow, Andrew Harp, Geoffrey Irving, Michael Isard,
  Yangqing Jia, Rafal Jozefowicz, Lukasz Kaiser, Manjunath Kudlur, Josh
  Levenberg, Dandelion Man\'{e}, Rajat Monga, Sherry Moore, Derek Murray, Chris
  Olah, Mike Schuster, Jonathon Shlens, Benoit Steiner, Ilya Sutskever, Kunal
  Talwar, Paul Tucker, Vincent Vanhoucke, Vijay Vasudevan, Fernanda Vi\'{e}gas,
  Oriol Vinyals, Pete Warden, Martin Wattenberg, Martin Wicke, Yuan Yu, and
  Xiaoqiang Zheng.
\newblock {TensorFlow}: Large-scale machine learning on heterogeneous systems,
  2015.
\newblock Software available from tensorflow.org.

\bibitem{pytorch}
Adam Paszke, Sam Gross, Soumith Chintala, Gregory Chanan, Edward Yang, Zachary
  DeVito, Zeming Lin, Alban Desmaison, Luca Antiga, and Adam Lerer.
\newblock Automatic differentiation in pytorch.
\newblock 2017.

\bibitem{keras}
Jesper~S{\"o}ren Dramsch and Contributors.
\newblock Complex-valued neural networks in keras with tensorflow, 2019.

\bibitem{theano}
{Theano Development Team}.
\newblock {Theano: A {Python} framework for fast computation of mathematical
  expressions}.
\newblock {\em arXiv e-prints}, abs/1605.02688, May 2016.

\bibitem{diffiqult}
Teresa Tamayo-Mendoza, Christoph Kreisbeck, Roland Lindh, and Al\'an
  Aspuru-Guzik.
\newblock Automatic differentiation in quantum chemistry with applications to
  fully variational hartree–fock.
\newblock {\em ACS central science}, 4(5):559--566, 2018.

\bibitem{pennylane}
Ville Bergholm, Josh Izaac, Maria Schuld, Christian Gogolin, M.~Sohaib Alam,
  Shahnawaz Ahmed, Juan~Miguel Arrazola, Carsten Blank, Alain Delgado, Soran
  Jahangiri, Keri McKiernan, Johannes~Jakob Meyer, Zeyue Niu, Antal Száva, and
  Nathan Killoran.
\newblock Pennylane: Automatic differentiation of hybrid quantum-classical
  computations.
\newblock {\em arXiv:1811.04968}, 2018.

\bibitem{dqc_pennylane}
Juan~Miguel Arrazola, Soran Jahangiri, Alain Delgado, Jack Ceroni, Josh Izaac,
  Antal Száva, Utkarsh Azad, Robert~A. Lang, Zeyue Niu, Olivia Di~Matteo,
  Romain Moyard, Jay Soni, Maria Schuld, Rodrigo~A. Vargas-Hernández, Teresa
  Tamayo-Mendoza, Cedric Yen-Yu Lin, Alán Aspuru-Guzik, and Nathan Killoran.
\newblock Differentiable quantum computational chemistry with pennylane.
\newblock {\em arXiv:2111.09967}, 2021.

\bibitem{dqc_kasim}
Muhammad~F Kasim, Susi Lehtola, and Sam~M Vinko.
\newblock Dqc: A python program package for differentiable quantum chemistry.
\newblock {\em The Journal of chemical physics}, 156(8):084801--084801, 2022.

\bibitem{ceder1998predicting}
Gerbrand Ceder.
\newblock Predicting properties from scratch.
\newblock {\em Science}, 280(5366):1099--1100, 1998.

\bibitem{Alchemy_bindingenergies}
Karthikeyan Saravanan, John~R. Kitchin, O.~Anatole von Lilienfeld, and John~A.
  Keith.
\newblock Alchemical predictions for computational catalysis: Potential and
  limitations.
\newblock {\em The Journal of Physical Chemistry Letters}, 8(20):5002--5007,
  2017.
\newblock PMID: 28938798.

\bibitem{2p+1}
X.~Gonze and J.-P. Vigneron.
\newblock Density-functional approach to nonlinear-response coefficients of
  solids.
\newblock {\em Phys. Rev. B}, 39:13120--13128, Jun 1989.

\bibitem{Mishkov2000}
Rumen~L. Mishkov.
\newblock Generalization of the formula of faa di bruno for a composite
  function with a vector argument.
\newblock {\em International Journal of Mathematics and Mathematical Sciences},
  24(7):481--491, 2000.

\bibitem{accurate_abinitio}
O~Anatole~von Lilienfeld.
\newblock Accurate ab initio energy gradients in chemical compound space.
\newblock {\em The Journal of chemical physics}, 131(16):164102--164102--6,
  2009.

\bibitem{HigherOrderAlchemicalDerivatives_SmithGunsteren1994}
P.~E. Smith and W.~F. van Gunsteren.
\newblock Predictions of free energy differences from a single simulation of
  the initial state.
\newblock {\em JCP}, 100:577, 1994.

\bibitem{lilienfeld_tuckerman}
O.~Anatole von Lilienfeld and Mark~E. Tuckerman.
\newblock Molecular grand-canonical ensemble density functional theory and
  exploration of chemical space.
\newblock {\em The Journal of Chemical Physics}, 125(15):154104, 2006.

\bibitem{pulay}
Peter Pulay.
\newblock Ab initio calculation of force constants and equilibrium geometries
  in polyatomic molecules. i. theory.
\newblock {\em Molecular physics}, 100(1):57--62, 2002.

\bibitem{giorgio_basisset}
Giorgio Domenichini, Guido~Falk von Rudorff, and O.~Anatole von Lilienfeld.
\newblock Effects of perturbation order and basis set on alchemical
  predictions.
\newblock {\em The Journal of chemical physics}, 153(14):144118--144118, 2020.

\bibitem{bse}
Benjamin~P. Pritchard, Doaa Altarawy, Brett Didier, Tara~D. Gibson, and
  Theresa~L. Windus.
\newblock New basis set exchange: An open, up-to-date resource for the
  molecular sciences community.
\newblock {\em Journal of Chemical Information and Modeling},
  59(11):4814--4820, Nov 2019.

\bibitem{chemical_space_balawender}
Robert {Balawender}, Michael {Lesiuk}, Frank {De Proft}, Christian {Van
  Alsenoy}, and Paul {Geerlings}.
\newblock {Exploring chemical space with alchemical derivatives: alchemical
  transformations of H through Ar and their ions as a proof of concept}.
\newblock {\em Physical Chemistry Chemical Physics (Incorporating Faraday
  Transactions)}, 21(43):23865--23879, November 2019.

\bibitem{jensen_basisset}
Maximilien~A Ambroise and Frank Jensen.
\newblock Probing basis set requirements for calculating core ionization and
  core excitation spectroscopy by the $delta$ self-consistent-field approach.
\newblock {\em Journal of chemical theory and computation}, 15(1):325--337,
  2019.

\bibitem{cohentannoudji}
Claude Cohen-Tannoudji, Bernard Diu, and Franck Laloë.
\newblock {\em Quantenmechanik, Band 2}.
\newblock Walter de Gruyter GmbH \& Co KG, 2008.

\bibitem{laguerre_polynomials}
Wolfram research, inc., wolfram|alpha knowledgebase, champaign, 2021.

\bibitem{dahl}
J.~P. Dahl and M.~Springborg.
\newblock The morse oscillator in position space, momentum space, and phase
  space.
\newblock {\em The Journal of chemical physics}, 88(7):4535--4547, 1988.

\bibitem{PySCF1}
Qiming Sun, Timothy~C. Berkelbach, Nick~S. Blunt, George~H. Booth, Sheng Guo,
  Zhendong Li, Junzi Liu, James~D. McClain, Elvira~R. Sayfutyarova, Sandeep
  Sharma, Sebastian Wouters, and Garnet Kin-Lic Chan.
\newblock Pyscf: the python-based simulations of chemistry framework.
\newblock {\em WIREs Computational Molecular Science}, 8(1):e1340, 2018.

\bibitem{PySCF2}
Qiming Sun, Xing Zhang, Samragni Banerjee, Peng Bao, Marc Barbry, Nick~S.
  Blunt, Nikolay~A. Bogdanov, George~H. Booth, Jia Chen, Zhi-Hao Cui, Janus~J.
  Eriksen, Yang Gao, Sheng Guo, Jan Hermann, Matthew~R. Hermes, Kevin Koh,
  Peter Koval, Susi Lehtola, Zhendong Li, Junzi Liu, Narbe Mardirossian,
  James~D. McClain, Mario Motta, Bastien Mussard, Hung~Q. Pham, Artem Pulkin,
  Wirawan Purwanto, Paul~J. Robinson, Enrico Ronca, Elvira~R. Sayfutyarova,
  Maximilian Scheurer, Henry~F. Schurkus, James E.~T. Smith, Chong Sun,
  Shi-Ning Sun, Shiv Upadhyay, Lucas~K. Wagner, Xiao Wang, Alec White,
  James~Daniel Whitfield, Mark~J. Williamson, Sebastian Wouters, Jun Yang,
  Jason~M. Yu, Tianyu Zhu, Timothy~C. Berkelbach, Sandeep Sharma, Alexander~Yu.
  Sokolov, and Garnet Kin-Lic Chan.
\newblock Recent developments in the pyscf program package.
\newblock {\em The Journal of Chemical Physics}, 153(2):024109, 2020.

\bibitem{numpy}
Charles~R. Harris, K.~Jarrod Millman, St{\'{e}}fan~J. van~der Walt, Ralf
  Gommers, Pauli Virtanen, David Cournapeau, Eric Wieser, Julian Taylor,
  Sebastian Berg, Nathaniel~J. Smith, Robert Kern, Matti Picus, Stephan Hoyer,
  Marten~H. van Kerkwijk, Matthew Brett, Allan Haldane, Jaime~Fern{\'{a}}ndez
  del R{\'{i}}o, Mark Wiebe, Pearu Peterson, Pierre G{\'{e}}rard-Marchant,
  Kevin Sheppard, Tyler Reddy, Warren Weckesser, Hameer Abbasi, Christoph
  Gohlke, and Travis~E. Oliphant.
\newblock Array programming with {NumPy}.
\newblock {\em Nature}, 585(7825):357--362, September 2020.

\bibitem{scipy}
Pauli Virtanen, Ralf Gommers, Travis~E. Oliphant, Matt Haberland, Tyler Reddy,
  David Cournapeau, Evgeni Burovski, Pearu Peterson, Warren Weckesser, Jonathan
  Bright, St{\'e}fan~J. {van der Walt}, Matthew Brett, Joshua Wilson, K.~Jarrod
  Millman, Nikolay Mayorov, Andrew R.~J. Nelson, Eric Jones, Robert Kern, Eric
  Larson, C~J Carey, {\.I}lhan Polat, Yu~Feng, Eric~W. Moore, Jake
  {VanderPlas}, Denis Laxalde, Josef Perktold, Robert Cimrman, Ian Henriksen,
  E.~A. Quintero, Charles~R. Harris, Anne~M. Archibald, Ant{\^o}nio~H. Ribeiro,
  Fabian Pedregosa, Paul {van Mulbregt}, and {SciPy 1.0 Contributors}.
\newblock {{SciPy} 1.0: Fundamental Algorithms for Scientific Computing in
  Python}.
\newblock {\em Nature Methods}, 17:261--272, 2020.

\bibitem{numba}
Siu~Kwan Lam, Antoine Pitrou, and Stanley Seibert.
\newblock Numba: A llvm-based python jit compiler.
\newblock In {\em Proceedings of the Second Workshop on the LLVM Compiler
  Infrastructure in HPC}, LLVM '15, New York, NY, USA, 2015. Association for
  Computing Machinery.

\bibitem{matplotlib}
J.~D. Hunter.
\newblock Matplotlib: A 2d graphics environment.
\newblock {\em Computing in Science \& Engineering}, 9(3):90--95, 2007.

\end{thebibliography}
\bibliographystyle{unsrt}


\clearpage
\onecolumngrid
\setcounter{section}{0}

\begin{center}
    \large \textbf{\papertitle}\\
    \large \textbf{ ---  Supplemental Material  ---}\\
    \vspace{\baselineskip}
    \normalsize Simon León Krug,$^{1,2}$ Guido Falk von Rudorff,$^{1,3}$ and O. Anatole von Lilienfeld$^{2,4,5}$\\
    \vspace{0.5\baselineskip}
    \small \textit{
    $^{1)}$University of Vienna, Computational Materials Physics, Kolingasse 14-16, 1090 Vienna, Austria\\
    $^{2)}$Machine Learning Group, Technische Universiät Berlin and Institute for the Foundations of Learning and Data,\\
    10587 Berlin, Germany\\
    $^{3)}$Institute for Pure and Applied Mathematics (IPAM), University of California,\\
    Los Angeles, 460 Portola Plaza, Los Angeles, CA~90095, USA\\
    $^{4)}$Vector Institute for Artificial Intelligence, Toronto, ON, M5S 1M1, Canada\\
    $^{5)}$Departments of Chemistry, Materials Science and Engineering, and Physics, University of Toronto, St. George Campus,\\
    Toronto, ON, Canada\\
    }
    \small (Dated: \today)
\end{center}

\section{Derivation of Eqs.~\ref{eq:Delta_E_coeff_final} and~\ref{eq:Delta_E_coeff_final_K_withTheta}}
\label{sec:derivation}
Starting with Eq.~\ref{eq:original_expansion}, we shift the index $p \longrightarrow p + 1$:
\begin{align}
	\Delta E &= \sum_{p = 1}^{\infty} \frac{1}{p!} \int_{\mathbb{R}^3} d\bm{r}\, \Delta v \frac{\partial^{p-1} \rho(\lambda, \bm{r})}{\partial \lambda^{p-1}} \bigg\vert_{\lambda = 0} = \sum_{p = 0}^{\infty} \frac{1}{(p+1)!} \int_{\mathbb{R}^3} d\bm{r}\, \Delta v \frac{\partial^{p} \rho(\lambda, \bm{r})}{\partial \lambda^{p}} \bigg\vert_{\lambda = 0}
	\label{eq:original_formula_appendix}
\end{align}
We want to reshape this expression by transferring the $\lambda$-dependency of the general $\rho (\lambda, \bm{r})$ to a parametrization of the coordinates $\bm{r} \rightarrow \bm{r}(\lambda)$ where $\bm{r}(\lambda = 0) =: \bm{r}_A$. We demand this parametrization to be analytic at $\lambda = 0$ and invertible around $\lambda = 0$, with the inverse denoted as $r^{-1} : \mathbb{R}^3 \longrightarrow [0,1]$. Both criteria are fulfilled by the parametrization given in Eq.~\ref{eq:r_lambda_full}. The Jacobian of this transform shall be denoted~$J$ with property $J\vert_{\lambda = 0} = 1$.

\begin{align}
    \Delta E &= \sum_{p = 0}^{\infty} \frac{1}{(p+1)!} \int_{\mathbb{R}^3} d\bm{r}(\lambda)\, J \,\Delta v(\bm{r}(\lambda)) \, \frac{\partial^{p} \rho(\lambda, \bm{r}(\lambda))}{\partial \lambda^{p}} \bigg\vert_{\lambda = 0}
\end{align}
Now we rewrite the electron density after the transformation, $\rho(\lambda, \bm{r}(\lambda))$, as a suitable new function~$\tilde{\rho}$:
\begin{align}
    \label{eq:rho_transform}
    \rho (\lambda, \bm{r}(\lambda)) &= \rho (r^{-1}(\bm{r}(\lambda)), \bm{r}(\lambda)) =: \tilde{\rho}(\bm{r}(\lambda))
\end{align}
\begin{align}
    \Rightarrow
    \label{eq:transformed_expansion}
    \Delta E &= \sum_{p = 0}^{\infty} \frac{1}{(p+1)!} \int_{\mathbb{R}^3} d\bm{r}(\lambda)\, J \,\Delta v(\bm{r}(\lambda)) \, \frac{\partial^{p} \tilde{\rho}(\bm{r}(\lambda))}{\partial \lambda^{p}} \bigg\vert_{\lambda = 0}
\end{align}

Use Faà di Bruno's formula for composite functions with a vector argument to re-express $\tilde{\rho}(\bm{r}(\lambda))$ as a chain of $\tilde{\rho}: \mathbb{R}^3 \longrightarrow \mathbb{R}^+ $ and $\bm{r}: [0,1] \longrightarrow \mathbb{R}^3$ (chain rule in 3 dimensions for the $p$-th derivative)~\cite{Mishkov2000}:
\begin{align}
	\label{eq:mishkow_raw}
	=& \sum_{p = 0}^{\infty} \frac{1}{(p+1)!}\int_{\mathbb{R}^3} d\bm{r}(\lambda)\,J\, \Delta v (\bm{r}(\lambda)) \sum_{0} \sum_{1} \cdots \sum_{p} \frac{p!}{\prod_{i = 1}^{p} \left( i! \right)^{k_i} \prod_{i = 1}^p \left( q_i^x ! \right) \left( q_i^y ! \right) \left( q_i^z ! \right)}\notag \\
	&\times \frac{\partial^k \tilde{\rho}(\bm{r}(\lambda))}{\partial x^{\mu_x}(\lambda) \partial y^{\mu_y}(\lambda) \partial z^{\mu_z}(\lambda)} \prod_{i = 1}^p \left( \frac{\partial^i x(\lambda)}{\partial \lambda^i} \right)^{q_i^x} \left( \frac{\partial^i y(\lambda)}{\partial \lambda^i} \right)^{q_i^y} \left( \frac{\partial^i z(\lambda)}{\partial \lambda^i} \right)^{q_i^z}\bigg\vert_{\lambda = 0}
\end{align}

\clearpage

\noindent
The sums run over all non-negative integer solutions of the following $p+1$ linear Diophantine equations:
\begin{align}
\label{eq:Diophantine_stairs}
	\sum_{0} : \qquad k_1 + 2k_2 + \cdots + pk_p &= p
\end{align}
\begin{align}
	\sum_{1} : \qquad q_1^x + q_1^y + q_1^z &= k_1 \\
	\vdots&\notag \\
	\sum_{p} : \qquad q_p^x + q_p^y + q_p^z &= k_p 
\end{align}
Furthermore, the conditions hold:
\begin{align}
	\mu_x &= q_1^x + \cdots + q_p^x\\
	\mu_y &= q_1^y + \cdots + q_p^y\\
	\mu_z &= q_1^z + \cdots + q_p^z\\
	\label{eq:k_to_mu}
	k &= \mu_x+\mu_y+\mu_z = k_1 + \cdots + k_p
\end{align}
Note that this only holds for the first and higher derivatives. In case of $p=0$, the sum $\sum_0$ shall be taken over all $p=\mu_x +\mu_y +\mu_z = 0$ which nicely reproduces the result of Eq.~\ref{eq:original_expansion} for $p=0$.

The chain of $\tilde{\rho}$ and $\bm{r}$ holds for general three-dimensional arguments -- it is not limited to Cartesian coordinates, i.e. $\bm{r} = (r, \vartheta, \phi)$ with $d\bm{r} = r^2 \sin(\vartheta) \, dr \,d\vartheta \,d\phi$ is equivalent.

Firstly, we can use the distributivity of summations when summing over different indices. Each $\sum_i$ only acts on its respective $k_i,q_i^x,q_i^y,q_i^z $, so the sums factorize; product and sums can be swapped:
\begin{align}
	\Delta E=& \sum_{p = 0}^{\infty} \frac{1}{p+1}\int_{\mathbb{R}^3} d\bm{r}(\lambda)\,J\, \Delta v (\bm{r}(\lambda)) \sum_{0} \frac{1}{\prod_{i = 1}^{p} \left( i! \right)^{k_i} k_i!}  \frac{\partial^k \tilde{\rho}(\bm{r}(\lambda))}{\partial x^{\mu_x}(\lambda) \partial y^{\mu_y}(\lambda) \partial z^{\mu_z}(\lambda)}\notag \\
	&\times \prod_{i = 1}^p \sum_{i} \frac{k_i!}{\left( q_i^x ! \right) \left( q_i^y ! \right) \left( q_i^z ! \right)} \left( \frac{\partial^i x(\lambda)}{\partial \lambda^i} \right)^{q_i^x} \left( \frac{\partial^i y(\lambda)}{\partial \lambda^i} \right)^{q_i^y} \left( \frac{\partial^i z(\lambda)}{\partial \lambda^i} \right)^{q_i^z}\bigg\vert_{\lambda = 0}
\end{align}
Secondly, we apply the multinomial theorem:
\begin{align}
	\Delta E=& \sum_{p = 0}^{\infty} \frac{1}{p+1} \sum_{0} \frac{1}{\prod_{i = 1}^{p} \left( i! \right)^{k_i} k_i!} \int_{\mathbb{R}^3} d\bm{r}(\lambda)\,J\, \Delta v (\bm{r}(\lambda)) \frac{\partial^k \tilde{\rho} (\bm{r}(\lambda))}{\partial x(\lambda)^{\mu_x} \partial y(\lambda)^{\mu_y} \partial z(\lambda)^{\mu_z}}\notag \\
	& \times \prod_{i = 1}^p \left( \frac{\partial^i x(\lambda)}{\partial \lambda^i} + \frac{\partial^i y(\lambda)}{\partial \lambda^i} + \frac{\partial^i z(\lambda)}{\partial \lambda^i} \right)^{k_i}\bigg\vert_{\lambda = 0}
\end{align}

\clearpage
\noindent
Thirdly, we perform $\mu_x,\mu_y,\mu_z$-times integration by parts. All limit terms equal zero because $\partial^{\mu_w} \rho_A/\partial w_A^{\mu_w} \longrightarrow 0$ as $w_A \longrightarrow \pm \infty$ for all ${\mu_w} \in \mathbb{N}_0$ and $w \in \lbrace x,y,z \rbrace$. Every finite system's electron density vanishes at infinite distance and so do its spatial derivatives. However, this holds only if $\partial^{\mu_w} \rho_A/\partial w_A^{\mu_w}$ decays faster than $\partial^{\mu_w} \Delta v/\partial w_A^{\mu_w}$ grows.
\begin{align}
\label{eq:Delta_E_with_lambda_der}
	\Delta E=& \sum_{p = 0}^{\infty} \frac{1}{p+1} \sum_{0} \frac{\left(-1\right)^k}{\prod_{i = 1}^{p} \left( i! \right)^{k_i} k_i!} \int_{\mathbb{R}^3} d\bm{r}(\lambda)\, \tilde{\rho}(\bm{r}(\lambda))\notag\\ &\times \frac{\partial^k}{\partial x(\lambda)^{\mu_x} \partial y(\lambda)^{\mu_y} \partial z(\lambda)^{\mu_z}} \left[J\, \Delta v (\bm{r}(\lambda))
	\prod_{i = 1}^p \left( \frac{\partial^i x(\lambda)}{\partial \lambda^i} + \frac{\partial^i y(\lambda)}{\partial \lambda^i} + \frac{\partial^i z(\lambda)}{\partial \lambda^i} \right)^{k_i}\right]\Bigg\vert_{\lambda = 0}
\end{align}
Within the brackets, only $\Delta v(\bm{r}(\lambda))$ is non-constant w.r.t. $x(\lambda),y(\lambda),z(\lambda)$ because:
\begin{align}
\frac{\partial}{\partial x} \prod_{i = 1}^p \left(  \frac{\partial^i x}{\partial \lambda^i} + \frac{\partial^i y}{\partial \lambda^i} + \frac{\partial^i z}{\partial \lambda^i} \right)^{k_i} = \prod_{i = 1}^p \left( \frac{\partial^i}{\partial \lambda^i} [x+y+z] \right)^{k_i} \left( \sum_{\substack{i = 1\\k_i \neq 0}}^p \frac{k_i \frac{\partial^i}{\partial \lambda^i} \frac{\partial}{\partial x}[x+y+z]}{\frac{\partial^i}{\partial \lambda^i} [x+y+z] } \right) = 0
\end{align}
As $\partial_x [x+y+z] = 1$, already the first derivative w.r.t $\lambda$ vanishes. The same goes for $y$ and $z$.
Furthermore, the Jacobian vanishes, too:
\begin{align}
    \frac{\partial}{\partial x} J = \frac{\partial}{\partial x} \text{det}(\bm{J}) = \frac{\partial}{\partial x} \text{det}  \left(\begin{array}{ccc}
\frac{\partial x}{\partial x_0} & \frac{\partial x}{\partial y_0} & \frac{\partial x}{\partial z_0} \\
    \frac{\partial y}{\partial x_0} & \frac{\partial y}{\partial y_0} & \frac{\partial y}{\partial z_0} \\
    \frac{\partial z}{\partial x_0} & \frac{\partial z}{\partial y_0} & \frac{\partial z}{\partial z_0}
\end{array}\right) = 0
\end{align}
Again, this holds for $y$ and $z$ as well.

Set $\lambda = 0$ wherever possible, revert the index shift from the beginning ($p \rightarrow p - 1$) and use $k = \mu_x + \mu_y + \mu_z$:
\begin{align}
\label{eq:Delta_E_coeff}
\Delta E=& \int_{\mathbb{R}^3} d\bm{r}_A\, \rho_A \left( \bm{r}_A \right) \, \, \mathcal{K} \left( \bm{r}_A, v_A, v_B \right) \\
\mathcal{K} \left( \bm{r}_A, v_A, v_B \right) = &\sum_{p = 1}^{\infty} \frac{1}{p} \sum_{S_p} \frac{\partial^{\mu_x + \mu_y + \mu_z} \Delta v (\bm{r}_A)}{\partial x_A^{\mu_x} \partial y_A^{\mu_y} \partial z_A^{\mu_z}} \left[\prod_{i = 1}^{p-1} \frac{ \left( \sum_{w \in \lbrace x,y,z \rbrace} \theta_{w,i} \right)^{k_i}}{k_i!} \right]\\
\theta_{w,i} := & - \frac{1}{i!} \frac{\partial^i w (\lambda)}{\partial \lambda^i} \bigg\vert_{\lambda = 0}\\
S_p := & \left\lbrace \mu_x, \mu_y, \mu_z, k_1, \dots, k_{p-1} \in \mathbb{N}_0 \, \Bigg\vert \, p-1 = \sum_{i=1}^{p-1} i \cdot k_i , \, \mu_x + \mu_y + \mu_z = \sum_{i=1}^{p-1} k_i\right\rbrace
\end{align}
\noindent
Eqs. \ref{eq:original_formula_appendix} and \ref{eq:Delta_E_coeff} match order-wise.

\clearpage
\section{Solutions to the linear Diophantine equations}
Below are a list of the first few $(3 + (p-1))$-tuples of $S_p$:

\begin{longtable}{c|l}
    $p$ &  $[ \mu_x, \mu_y, \mu_z, k_1, \dots, k_{p-1}]$\\
     \hline
     1& [0,0,0]\\
     \hline
     2 & [1,0,0,1],[0,1,0,1],[0,0,1,1] \\
     \hline
     3 & [0,0,1,0,1],[0,0,2,2,0],[0,1,0,0,1],[0,1,1,2,0],[0,2,0,2,0], [1,0,0,0,1],[1,0,1,2,0],[1,1,0,2,0],[2,0,0,2,0]\\
     \hline
     4 & [0,0,1,0,0,1],[0,0,2,1,1,0],[0,0,3,3,0,0],[0,1,0,0,0,1],[0,1,1,1,1,0],[0,1,2,3,0,0],[0,2,0,1,1,0],[0,2,1,3,0,0],\\
       & [0,3,0,3,0,0],[1,0,0,0,0,1],[1,0,1,1,1,0],[1,0,2,3,0,0],[1,1,0,1,1,0],[1,1,1,3,0,0],[1,2,0,3,0,0],[2,0,0,1,1,0],\\
       & [2,0,1,3,0,0],[2,1,0,3,0,0],[3,0,0,3,0,0]\\
     \hline
     5 & [0,0,1,0,0,0,1],[0,0,2,0,2,0,0],[0,0,2,1,0,1,0],[0,0,3,2,1,0,0],[0,0,4,4,0,0,0],[0,1,0,0,0,0,1],[0,1,1,0,2,0,0],\\
     & [0,1,1,1,0,1,0],[0,1,2,2,1,0,0],[0,1,3,4,0,0,0],[0,2,0,0,2,0,0],[0,2,0,1,0,1,0],[0,2,1,2,1,0,0],[0,2,2,4,0,0,0],\\
     & [0,3,0,2,1,0,0],[0,3,1,4,0,0,0],[0,4,0,4,0,0,0],[1,0,0,0,0,0,1],[1,0,1,0,2,0,0],[1,0,1,1,0,1,0],[1,0,2,2,1,0,0],\\
     & [1,0,3,4,0,0,0],[1,1,0,0,2,0,0],[1,1,0,1,0,1,0],[1,1,1,2,1,0,0],[1,1,2,4,0,0,0],[1,2,0,2,1,0,0],[1,2,1,4,0,0,0],\\
     & [1,3,0,4,0,0,0],[2,0,0,0,2,0,0],[2,0,0,1,0,1,0],[2,0,1,2,1,0,0],[2,0,2,4,0,0,0],[2,1,0,2,1,0,0],[2,1,1,4,0,0,0],\\
     & [2,2,0,4,0,0,0],[3,0,0,2,1,0,0],[3,0,1,4,0,0,0],[3,1,0,4,0,0,0],[4,0,0,4,0,0,0]\\
     \hline
     6 & [0,0,1,0,0,0,0,1],[0,0,2,0,1,1,0,0],[0,0,2,1,0,0,1,0],[0,0,3,1,2,0,0,0],[0,0,3,2,0,1,0,0],[0,0,4,3,1,0,0,0],\\
     & [0,0,5,5,0,0,0,0],[0,1,0,0,0,0,0,1],[0,1,1,0,1,1,0,0],[0,1,1,1,0,0,1,0],[0,1,2,1,2,0,0,0],[0,1,2,2,0,1,0,0],\\
     & [0,1,3,3,1,0,0,0],[0,1,4,5,0,0,0,0],[0,2,0,0,1,1,0,0],[0,2,0,1,0,0,1,0],[0,2,1,1,2,0,0,0],[0,2,1,2,0,1,0,0],\\
     & [0,2,2,3,1,0,0,0],[0,2,3,5,0,0,0,0],[0,3,0,1,2,0,0,0],[0,3,0,2,0,1,0,0],[0,3,1,3,1,0,0,0],[0,3,2,5,0,0,0,0],\\
     & [0,4,0,3,1,0,0,0],[0,4,1,5,0,0,0,0],[0,5,0,5,0,0,0,0],[1,0,0,0,0,0,0,1],[1,0,1,0,1,1,0,0],[1,0,1,1,0,0,1,0],\\
     & [1,0,2,1,2,0,0,0],[1,0,2,2,0,1,0,0],[1,0,3,3,1,0,0,0],[1,0,4,5,0,0,0,0],[1,1,0,0,1,1,0,0],[1,1,0,1,0,0,1,0],\\
     & [1,1,1,1,2,0,0,0],[1,1,1,2,0,1,0,0],[1,1,2,3,1,0,0,0],[1,1,3,5,0,0,0,0],[1,2,0,1,2,0,0,0],[1,2,0,2,0,1,0,0],\\
     & [1,2,1,3,1,0,0,0],[1,2,2,5,0,0,0,0],[1,3,0,3,1,0,0,0],[1,3,1,5,0,0,0,0],[1,4,0,5,0,0,0,0],[2,0,0,0,1,1,0,0],\\
     & [2,0,0,1,0,0,1,0],[2,0,1,1,2,0,0,0],[2,0,1,2,0,1,0,0],[2,0,2,3,1,0,0,0],[2,0,3,5,0,0,0,0],[2,1,0,1,2,0,0,0],\\
     & [2,1,0,2,0,1,0,0],[2,1,1,3,1,0,0,0],[2,1,2,5,0,0,0,0],[2,2,0,3,1,0,0,0],[2,2,1,5,0,0,0,0],[2,3,0,5,0,0,0,0],\\
     & [3,0,0,1,2,0,0,0],[3,0,0,2,0,1,0,0],[3,0,1,3,1,0,0,0],[3,0,2,5,0,0,0,0],[3,1,0,3,1,0,0,0],[3,1,1,5,0,0,0,0],\\
     & [3,2,0,5,0,0,0,0],[4,0,0,3,1,0,0,0],[4,0,1,5,0,0,0,0],[4,1,0,5,0,0,0,0],[5,0,0,5,0,0,0,0]\\
     \hline
     \dots & \dots \\
\end{longtable}

\section{Electron density of hydrogen-like atom}
\label{sec:derivation_bar_rho_A}
The solutions of the hydrogen-like atom with nuclear charge $Z_A$ in spherical coordinates and atomic units are~\cite{cohentannoudji}:
\begin{align}
    \Psi_{nlm} (r, \vartheta, \phi) &= R_{nl}\left( r \right) Y_{lm} \left(\vartheta, \phi \right)\\
    R_{nl}\left( r \right) &= - \left( \frac{2Z_A}{n} \right)^{3/2} \left( \frac{2Z_A r}{n} \right)^{l}  L_{n-l-1}^{(2l+1)}\left( \frac{2Z_Ar}{n} \right) \sqrt{\frac{(n-l-1)!}{2n (n+l)!}} \, \exp{\left(- \frac{Z_Ar}{n}\right)} \quad \text{ ,}
\end{align}
with the radial contribution $R_{nl}$, generalized Laguerre polynomials $L$ and the spherical harmonics $Y_{lm}$. We neglected the change of the reduced mass $\mu$ with increasing nuclear mass and chose $\mu \approx m_e$, hence $a_{\mu} \approx a_0 =1$. The spherically averaged electron density $\bar{\rho}_A$, depending only on $r_A$, $n$ and $Z_A$, is then given by:
\begin{align}
    \bar{\rho}_A (r_A, n, Z_A) &= \frac{1}{n^2} \sum_{l,m} \bra{\Psi^*_{nlm} (r, \vartheta, \phi)} \delta (\bm{r} - \bm{r}_A) \ket{\Psi_{nlm} (r, \vartheta, \phi)} \\
    &= \frac{1}{n^2} \sum_{l = 0}^{n-1} \frac{2l+1}{4\pi} \left( \frac{2Z_A}{n} \right)^3 \left( \frac{2Z_A r_A}{n} \right)^{2l} \left( L_{n-l-1}^{(2l+1)}\left( \frac{2Z_Ar_A}{n}\right) \right)^2 \frac{(n-l-1)!}{2n (n+l)!} \exp{\left(- \frac{2Z_Ar_A}{n}\right)}
\end{align}

\section{Derivation of Eq.~\ref{eq:Delta_E_coeff_radial}}
\label{sec:derivation_rad}
A simplification for radially symmetric systems can be derived from Eqs.~\ref{eq:Delta_E_coeff_final} and~\ref{eq:Delta_E_coeff_final_K_withTheta}; instead of $x_A, y_A, z_A$, we choose spherical coordinates $r_A, \vartheta_A, \phi_A$. Since any mono-atomic system is independent of its angles, $\vartheta_A, \phi_A$ can be dropped in the application of Faà di Bruno's formula in Eq.~\ref{eq:mishkow_raw} and the angular integration $\int d{\Omega}$ reduces to a constant of $4\pi$. Additionally, the sum over the set $S_p$ reduces to $T_p$ below, as two of three $\mu$'s are zero:
\begin{align}
	\Delta E_{\text{AIT}}=& \int\limits_{0}^{\infty} dr_A \, r_A^2 \int d\Omega \,\, \rho_A \left( r_A, n, Z_A \right) \sum_{p = 1}^{\infty} \frac{\left(1 - \frac{Z_B}{Z_A}\right)^{p-1}}{p} \sum_{T_p} \left[ \frac{\partial^{\mu_r}}{\partial r_A^{\mu_r}} \frac{-Z_B + Z_A}{r_A} \right] \cdot \left[ \prod_{i = 1}^{p-1} \frac{ r_A^{k_i}}{k_i!} \right]\\
	\label{eq:gp}
	=& \int\limits_0^{\infty} dr_A \, 4\pi r_A \, \bar{\rho}_A(r_A, n, Z_A) \sum_{p = 1}^{\infty}  \underbrace{\left( \frac{1}{p}\sum_{T_p} \frac{(-1)^{\mu_r} \cdot \mu_r!}{\prod_i^{p-1} k_i!} \right)}_{=: g_p} \, (-Z_B + Z_A) \left( 1 - \frac{Z_B}{Z_A} \right)^{p-1} \\
	\label{eq:Tp}
    T_p :=& \left\lbrace \mu_r, k_1, \dots, k_{p-1} \in \mathbb{N}_0 \, \Bigg\vert \, p-1 = \sum_{i=1}^{p-1} i \cdot k_i , \, \mu_r = \sum_{i=1}^{p-1} k_i\right\rbrace
\end{align}

The first few $g_p$ are:
\begin{table}[H]
\centering
\begin{tabular}{ c|ccccccc } 
    $p$ &  1 & 2&3&4&5&6&\dots\\
    \hline
    $g_p$ & $1$ & $-\frac{1}{2}$& $-\frac{1}{6}$ & $-\frac{1}{24}$ & $+\frac{1}{120}$ & $+\frac{19}{760}$&\dots\\
\end{tabular}
\end{table}

\section{Proof of Eq.~\ref{eq:gausslaw}}
\label{sec:derivation_gausslaw}

\begin{figure}[t]
    \centering
    \includegraphics[width=0.45\textwidth]{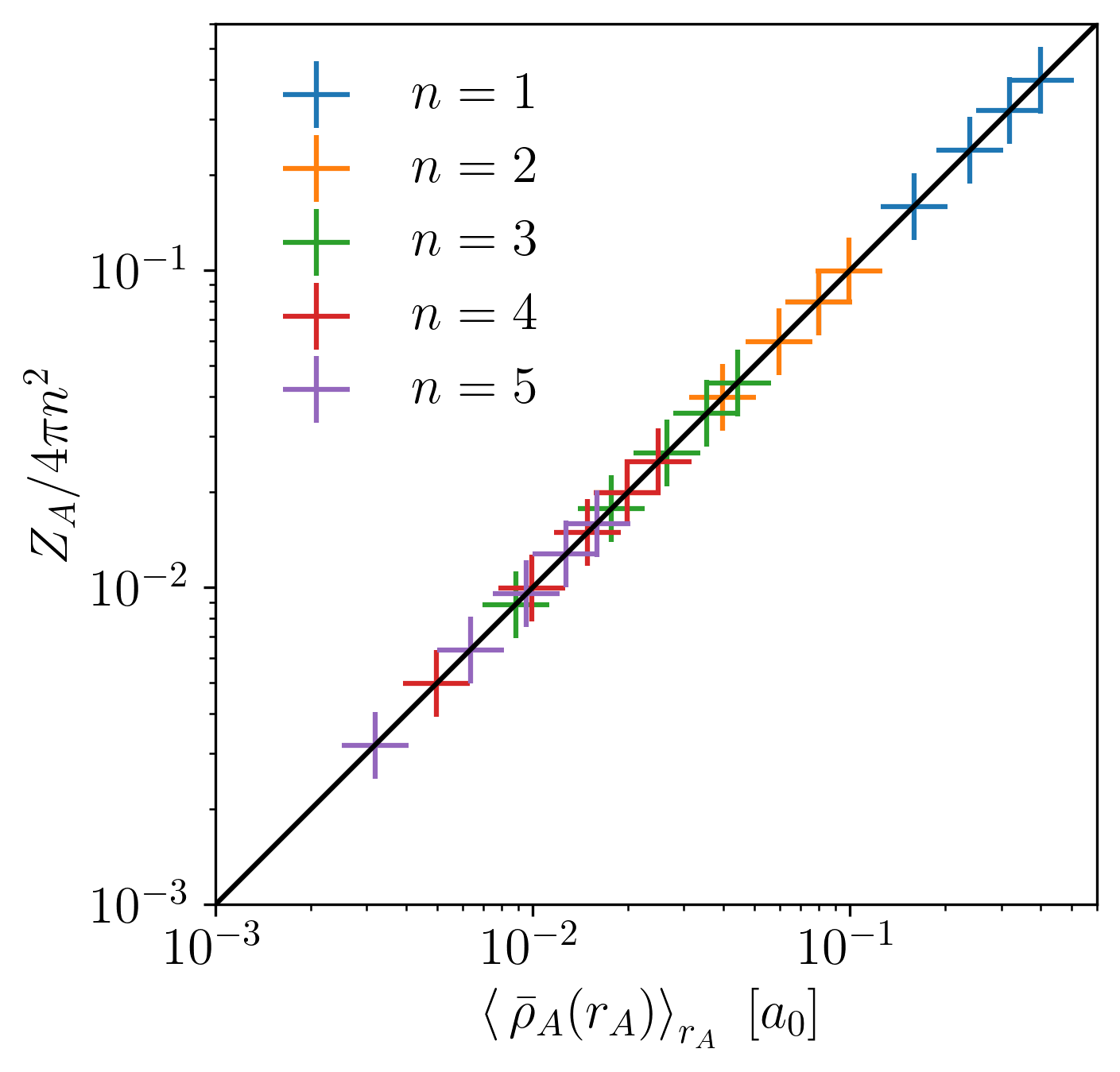}
    \caption{The hydrogen-like atom: $Z_A/4\pi n^2$ versus radial expectation value of the electron density $\left< \, \bar{\rho}_A(r_A) \right>_{r_A}$ at different principal quantum numbers $n$ for $Z_A \in \lbrace 1,2,3,4,5 \rbrace$. The numbers match up to $10^{-16}$ Ha}
    \label{fig:radial_Z_A_law}
\end{figure}

Eq.~\ref{eq:gausslaw} is visualized in Fig. \ref{fig:radial_Z_A_law}. However, it can also be proven with the following orthogonality relation of the Laguerre polynomials~\cite{laguerre_polynomials}:
\begin{align}
    \label{eq:laguerre_orthogonality}
    \int\limits_0^{\infty} dx \, \, e^{-x} \, x^k \, L^{(k)}_{(m)}\left( x \right) \, L^{(k)}_{(n)}\left( x \right) = \frac{(n+k)!}{n!} \delta_{mn}
\end{align}
Thus, we can write:
\begin{align}
    \frac{Z_A}{4\pi n^2} =& \int\limits_0^{\infty} dr_A \, r_A \, \bar{\rho}_A(r_A, n, Z_A)\\
    =& \,\frac{1}{n^2} \sum_{l = 0}^{n-1} \frac{2l+1}{4\pi} \left( \frac{2Z_A}{n} \right)^3 \frac{(n-l-1)!}{2n (n+l)!} \int\limits_0^{\infty} dr_A \, r_A \cdot \left( \frac{2Z_A r_A}{n} \right)^{2l} \left( L_{n-l-1}^{(2l+1)}\left( \frac{2Z_Ar_A}{n}\right) \right)^2 \exp{\left(- \frac{2Z_Ar_A}{n}\right)} \\
    \intertext{Substitute $\nu = 2Z_Ar_A/n$:}
    = & \,\frac{1}{n^2} \sum_{l = 0}^{n-1} \frac{Z_A }{4\pi n^2} (2l+1) \frac{(n-l-1)!}{(n+l)!} \int\limits_0^{\infty} d\nu \, \, e^{-\nu} \, \nu^{2l+1} \, \left( L_{n-l-1}^{(2l+1)}\left( \nu \right) \right)^2\\
    \notag
    = & \,\frac{Z_A}{4\pi n^2} \frac{1}{n^2} \sum_{l = 0}^{n-1} (2l+1) = \,\frac{Z_A}{4\pi n^2}
\end{align}

\section{The Dirac delta potential}
Consider a particle in the one-dimensional potential of a Dirac delta function
\begin{align}
    v(x) &= -b \cdot \delta (x) \qquad b > 0 \quad \text{,}
\end{align}
with energy eigenvalue and wave functions
\begin{align}
    E &= - \frac{b^2}{2} \label{eq:Diracdelta_energy} \\
    \Psi (x) &= \sqrt{b} \cdot \exp (- b |x|) \quad \text{.}
\end{align}
\noindent
Using AIT to obtain the energy difference between two such systems A and B with well depths $b_A$ and $b_B$:
\begin{align}
    \Delta E_{BA} &= \int\limits_{-\infty}^{\infty} dx_A \, \rho_A(x_A) \, \, \mathcal{K} \left( x_A, v_A, v_B \right)\\
    &= \int\limits_{-\infty}^{\infty} dx_A \, b_A \cdot \exp \left(-2b_A |x_A| \, \right) \, \sum_{p=1}^{\infty} \frac{\left(1 - \frac{b_B}{b_A} \right)^{p-1}}{p} \sum_{T_p} \, \left(b_A - b_B\right) \frac{\partial^{\mu_x} \delta(x_A)}{\partial x_A^{\mu_x}} \cdot \left[ \prod_{i = 1}^{p-1} \frac{ x_A^{k_i}}{k_i!} \right]\\
    &= b_A^2 \, \sum_{p=1}^{\infty} \frac{\left(1 - \frac{b_B}{b_A} \right)^{p}}{p} \sum_{T_p} \, \left[ \prod_{i = 1}^{p-1} \frac{ 1}{k_i!} \right] \, \int\limits_{-\infty}^{\infty} dx_A \, \exp \left(-2b_A |x_A| \, \right) \, x_A^{\mu_x}\, \frac{\partial^{\mu_x} \delta(x_A)}{\partial x_A^{\mu_x}}
\end{align}
with $T_p$ same as in Eq.~\ref{eq:Tp}.

The integral can be evaluated by applying the distributional derivative of the Dirac delta function:
\begin{align}
   &= b_A^2 \, \sum_{p=1}^{\infty} \frac{\left(1 - \frac{b_B}{b_A} \right)^{p}}{p} \sum_{T_p} \, \left[ \prod_{i = 1}^{p-1} \frac{ 1}{k_i!} \right] \, \mu_x ! \, (-1)^{\mu_x} \\
    &= b_A \, (b_A - b_B) \sum_{p=1}^{\infty} g_p \left( 1 - \frac{b_B}{b_A}\right)^{p-1}
\end{align}
with $g_p$ same as in Eq.~\ref{eq:gp}. Similar to the hydrogen-like atom, one can numerically show:
\begin{align}
    \sum_{p=1}^{\infty} g_p \, \xi^{p-1} = 1 - \frac{\xi}{2} \qquad  \xi \in \mathbb{R}
\end{align}
Finally, we obtain the energy difference from AIT:
\begin{align}
    \Delta E_{BA} = - \frac{b_B^2 - b_A^2}{2}
\end{align}
This is identical to the energy difference calculated from Eq.~\ref{eq:Diracdelta_energy}.

\section{The quantum harmonic oscillator}
Consider the potential of the one-dimensional harmonic oscillator
\begin{align}
    v(x) &= \frac{\omega^2}{2}x^2 \qquad \omega > 0 \quad \text{,}
\end{align}
with energy eigenvalue and wave functions
\begin{align}
    E_n &= \omega \, (n+\frac{1}{2}) \label{eq:harmosc_energy} \\
    \Psi_n (x) &= \frac{1}{\sqrt{2^n \, n!}} \left( \frac{\omega}{\pi} \right)^{1/4} \exp \left(-\frac{\omega x^2}{2} \right) \, H_n \left( \sqrt{\omega} x\right)\quad \text{.}
\end{align}
where $H_n$ are the physicist's Hermite polynomials \cite{cohentannoudji}.

Using AIT to obtain the energy difference between two such systems A and B with frequencies $\omega_A$ and $\omega_B$ proves to be difficult analytically, as well as numerically. However, the numerical difficulties come from the convergence behavior of the series in $\mathcal{K}(x, v_B, v_A)$ and can be evaded by adding a regulatory energy constant $\Lambda_{\text{reg}} \gg \Delta E_{BA}$ to initial and final potential. The energy difference between the systems $\Delta E_{BA}$ and the wave function are unaffected by this but the convergence behavior of the AIT kernel changes towards more favorable regimes.

With this, Eqs.~\ref{eq:Delta_E_harmosc_analyt} and~\ref{eq:Delta_E_harmosc_AIT} equal one another with small numerical error as seen in Fig.~\ref{fig:harmosc_proof}.
\begin{align}
    \label{eq:Delta_E_harmosc_analyt}
    \Delta E_{BA} &= (\omega_B - \omega_A) (n+\frac{1}{2})\\
    \label{eq:Delta_E_harmosc_AIT}
    \Delta E_{BA} &=  \int\limits_{-\infty}^{+\infty} dx_A \, \rho_A(x_A) \, \, \mathcal{K} \left( x_A, v_A + \Lambda_{\text{reg}} , v_B + \Lambda_{\text{reg}} \right)
\end{align}

\section{The Morse potential}
Consider the one-dimensional Morse potential centered around $x_0$ with well depth $D$ and range parameter $a$ \cite{dahl}
\begin{align}
    v(x) &= D \cdot \left( \exp (-2a \cdot (x - x_0)) - 2\exp (-a \cdot (x - x_0))\right) \qquad D,a > 0 \quad \text{,}
\end{align}
with energy eigenvalue and wave functions
\begin{align}
    E_n &= \sqrt{2D} \, a\cdot \left(n+\frac{1}{2}\right) \cdot \left(1 - \frac{a}{2\sqrt{2D}}\left(n+\frac{1}{2}\right) \right) - D \label{eq:morse_energy} \\
    \Psi_n (x) &= N(z,n) \sqrt{a} \, \xi^{z-n-1/2} e^{-\xi/2} L^{(2z-2n-1)}_n (\xi)\\
    z &= \frac{2D}{a}\\
    \xi &= 2z\cdot e^{-a(x-x_0)}\\
    N(z,n) &= \sqrt{\frac{(2z-2n-1) \, \Gamma (n+1)}{\Gamma (2z-n)}} \quad \text{,}
\end{align}
where $L$ are the generalized Laguerre polynomials.

\begin{figure}[h!]
    \centering
    \includegraphics[width=0.96\textwidth]{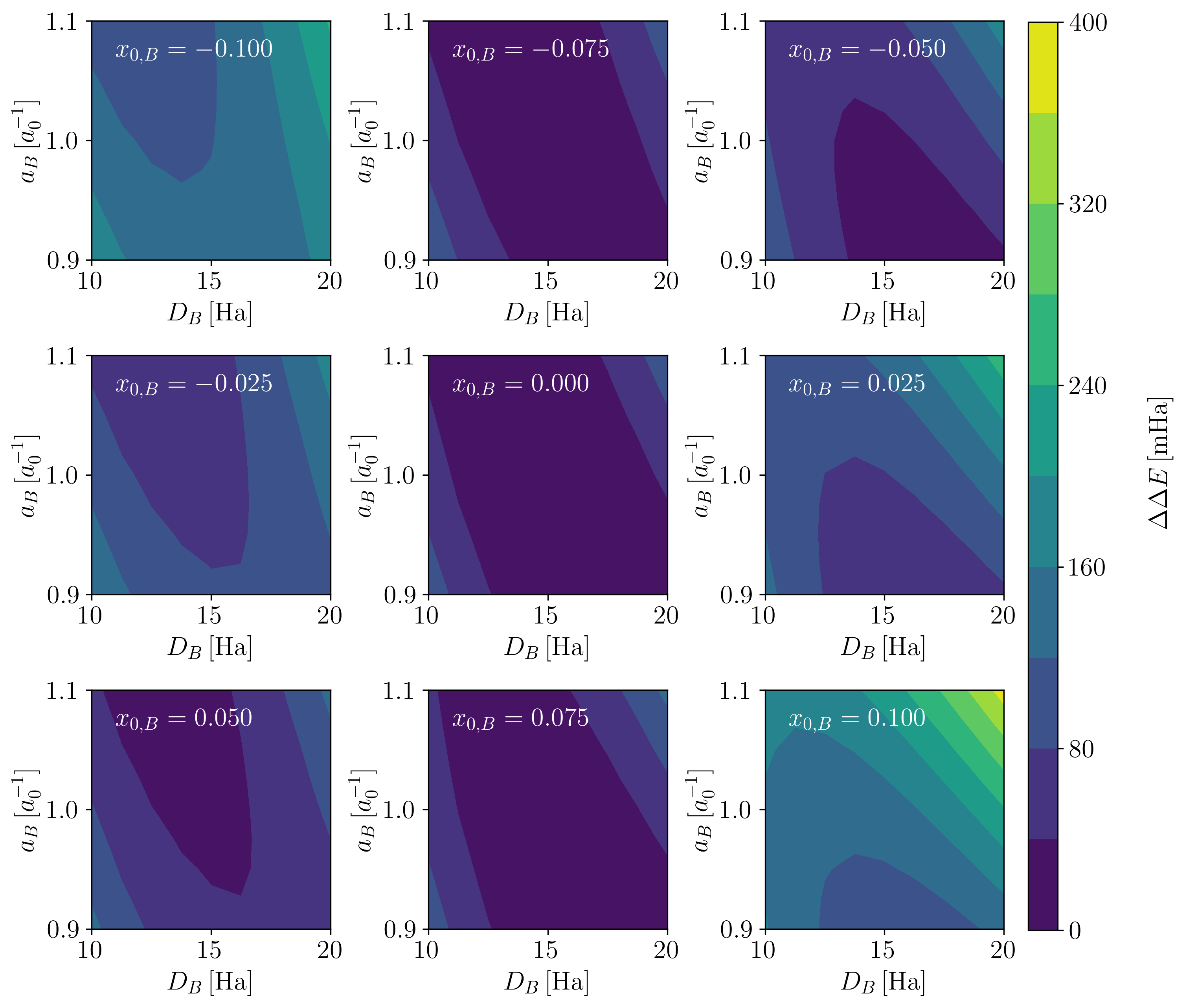}
    \caption{AIT for the three parameters depth $D$, width $a$ and minimum position $x_0$ of the Morse potential: the error $\Delta \Delta E = |\Delta E_{\text{exact}} - \Delta E_{\text{AIT}}|$ between the analytically known energy $\Delta E_{\text{exact}}$ and the AIT energy $\Delta E_{\text{AIT}}$ up to and including fifth perturbation order $p$ for quantum number $n=0$. The initial system's parameter are $D_A = 15$, $a_A = 1$, $x_{0,A} = 0$ in atomic units.}
    \label{fig:morse_3d_proof}
\end{figure}

\begin{figure}[h!]
    \subfloat[]{
        \centering
        \includegraphics[width=0.49\textwidth]{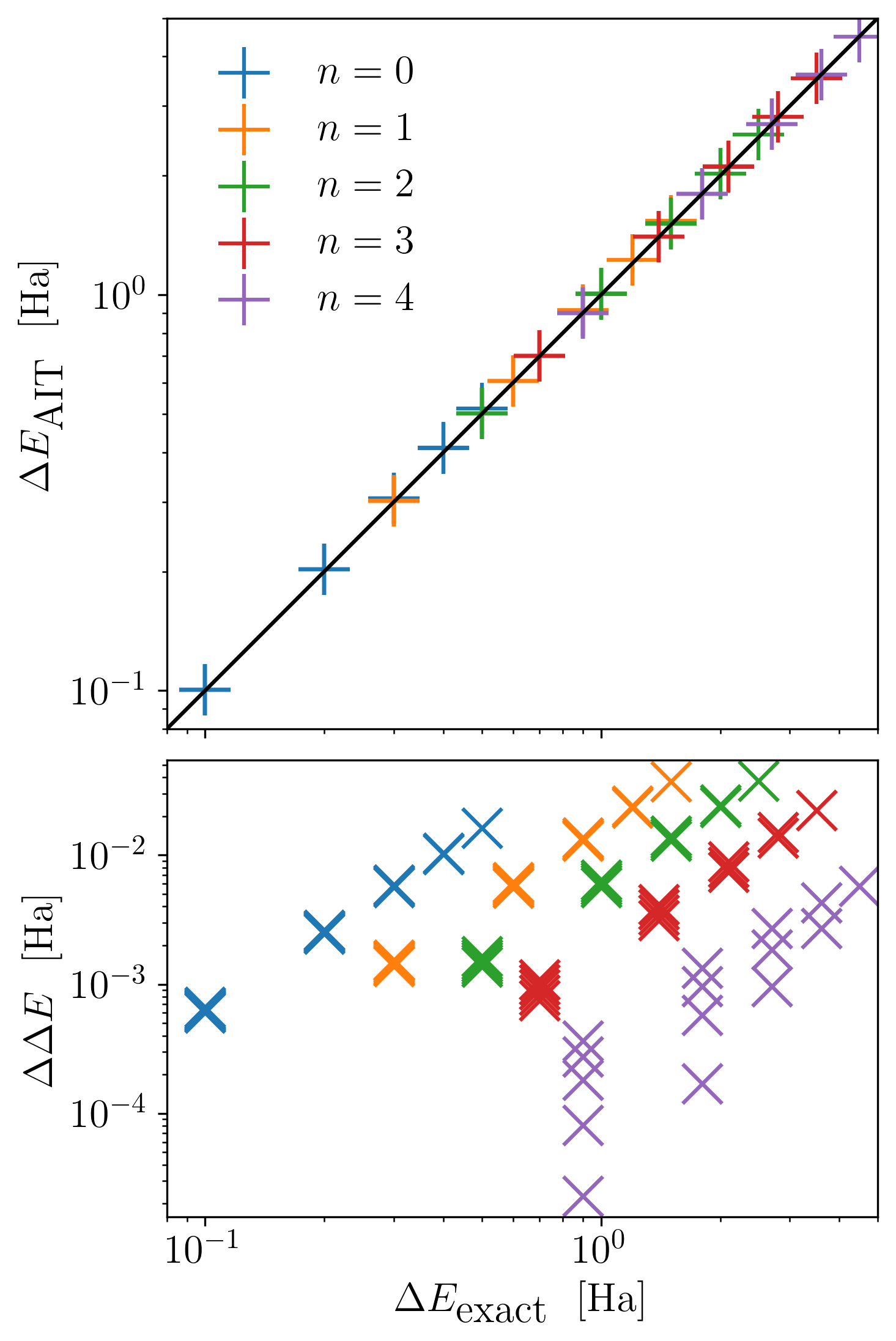}%
        \label{fig:harmosc_proof}%
    }
    \subfloat[]{
        \centering
        \includegraphics[width=0.49\textwidth]{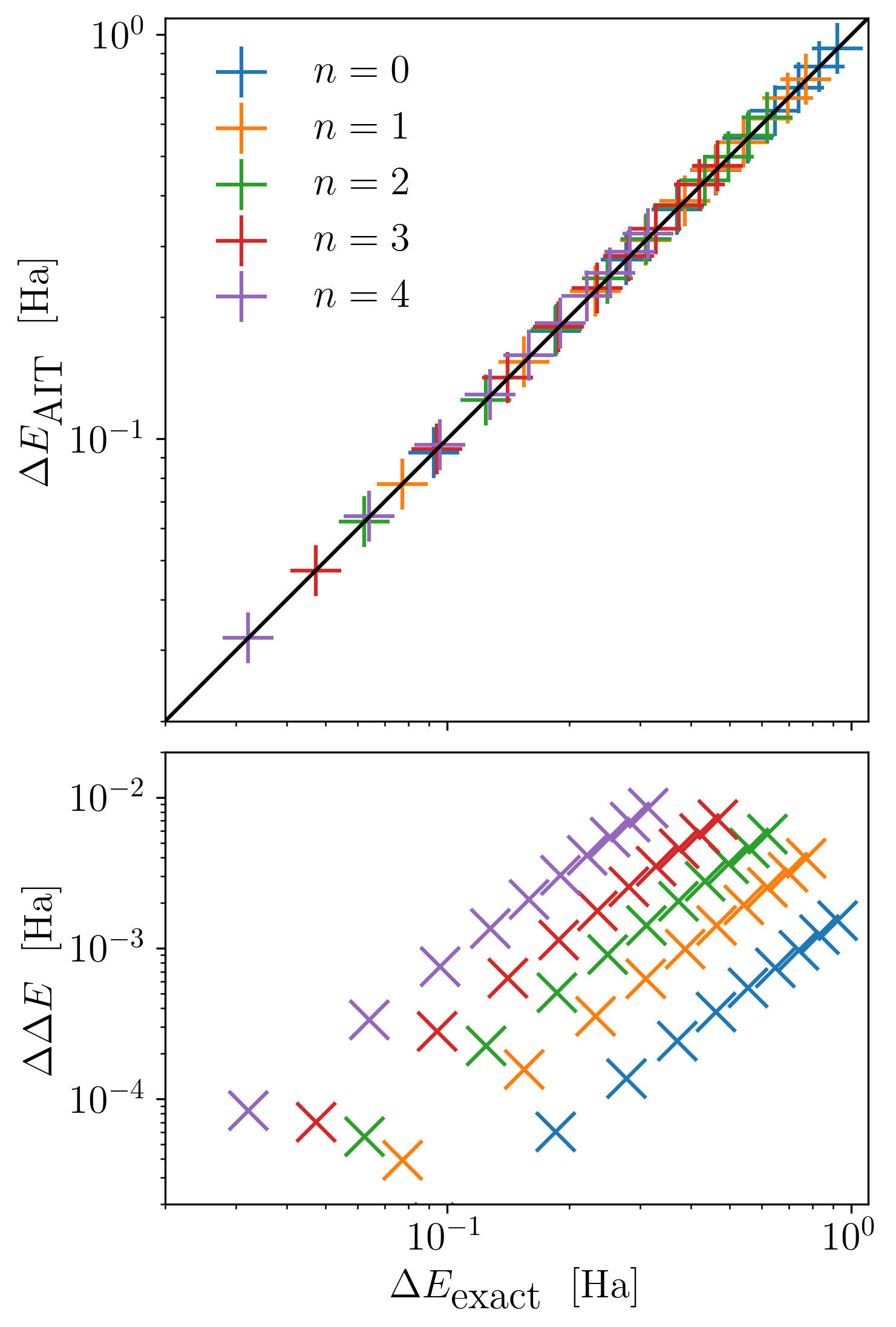}%
        \label{fig:morse_proof}%
    }
    \caption{AIT for two textbook examples of potentials: the analytically known energy $\Delta E_{\text{exact}}$ scattered against the AIT energy $\Delta E_{\text{AIT}}$ up to and including fifth perturbation order $p$, below the error $\Delta \Delta E = |\Delta E_{\text{exact}} - \Delta E_{\text{AIT}}|$, for different quantum numbers $n$. In~\ref{fig:harmosc_proof}, the quantum harmonic oscillator with $\omega_A, \omega_B \in \lbrace 11.0, 11.2, \dots , 12.0 \rbrace$ and $\omega_A < \omega_B$. In~\ref{fig:morse_proof}, the Morse potential centered at $x_0 = 0$ with $a_A = a_B = 1$, $D_A = 22$, $D_B = D_A - dZ$ and $dZ \in \lbrace 0.1, 0.2, \dots, 1.0 \rbrace$.}
    \label{fig:morse_harmosc_proof}
\end{figure}

Again, adding a regulatory constant $\Lambda_{\text{reg}}$ to initial and final potential in the AIT kernel enables us to obtain the energy difference $\Delta E_{BA}$ between two systems $A$ and $B$ with small numerical error as seen in Figs.~\ref{fig:morse_3d_proof} and~\ref{fig:morse_proof}.

\clearpage

\section{Periodic potentials in the Alchemical Integral transform}
Consider the AIT in $n$ dimensions with periodic initial and final potentials $v_A, v_B: \mathbb{R}^n \rightarrow \mathbb{R}$ of $N_1 N_2 \dots  N_n$ many cells and cell vector $\bm{R}$ with length $L_i = N_i R_i $, such that both can be written as $n$ sums of a single potential $v^{\text{cell}}$:
\begin{align}
    v_{A,B} (\bm{r}) &= \sum_{j_1 = 0}^{N_1-1} \cdots \sum_{j_n = 0}^{N_n-1} v^{\text{cell}}(r_1 + j_1 \cdot R_1, \dots, r_n + j_n \cdot R_n)
\end{align}
    As $L_1, \dots , L_n \rightarrow \infty$, any central cell experiences only an effective potential $v^{\text{eff}}_{A,B} (\bm{r})$. As all $N_1 N_2 \dots  N_n$ cells can be treated identically in the limit, one needs to evaluate the integral over $\mathbb{R}^n$ only for one central cell $\Omega^n = [0,R_1) \times \dots \times [0,R_n)$, then multiply by the number of cells:
\begin{align}
    \Delta E_{BA} &= \int_{\mathbb{R}^n} d\bm{r}_A\, \rho_A \left( \bm{r}_A \right) \, \, \mathcal{K} \left( \bm{r}_A, v_A, v_B \right)\\
    \Leftrightarrow \qquad \left[N_1 N_2 \dots N_n \right] \, \Delta E^{\text{cell}}_{BA} &= \left[N_1 N_2 \dots N_n \right] \,\int_{\Omega^n} d\bm{r}_A\, \rho_A \left( \bm{r}_A \right) \, \, \mathcal{K} \left( \bm{r}_A, v^{\text{eff}}_A, v^{\text{eff}}_B \right)
\end{align}
\noindent
Dropping the factors of $N_1 N_2 \dots  N_n$ on both sides gives access to the energy difference per cell $E^{\text{cell}}_{BA}$ and allows the use of AIT for periodic potentials.

\clearpage

\section{Noble gas or corresponding element as basis functions?}

\begin{figure}[h!]
    \subfloat[]{
        \centering
        \includegraphics[width=0.49\textwidth]{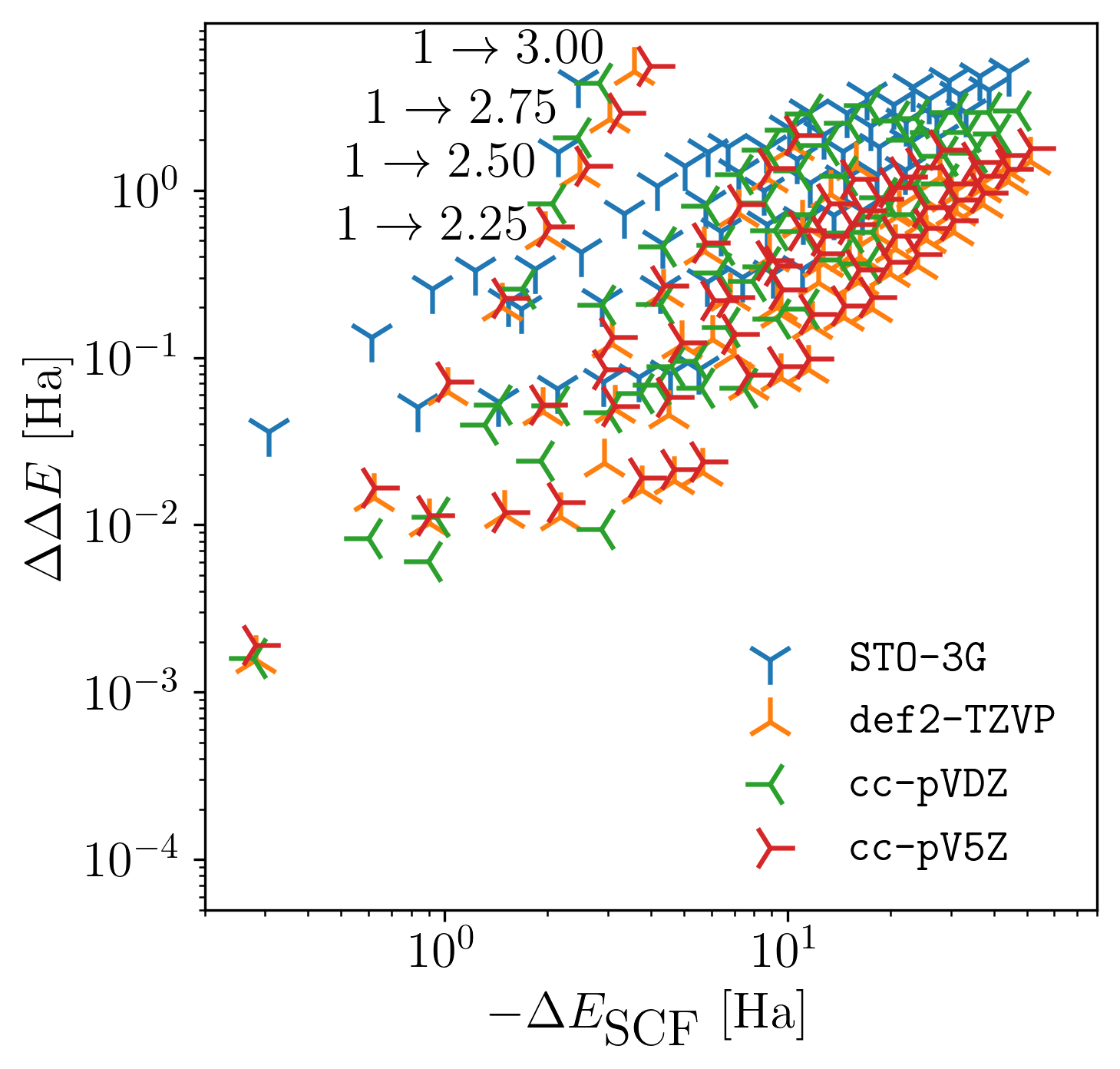}%
        \label{fig:monoatomics_proof_ind}%
    }
    \subfloat[]{
        \centering
        \includegraphics[width=0.49\textwidth]{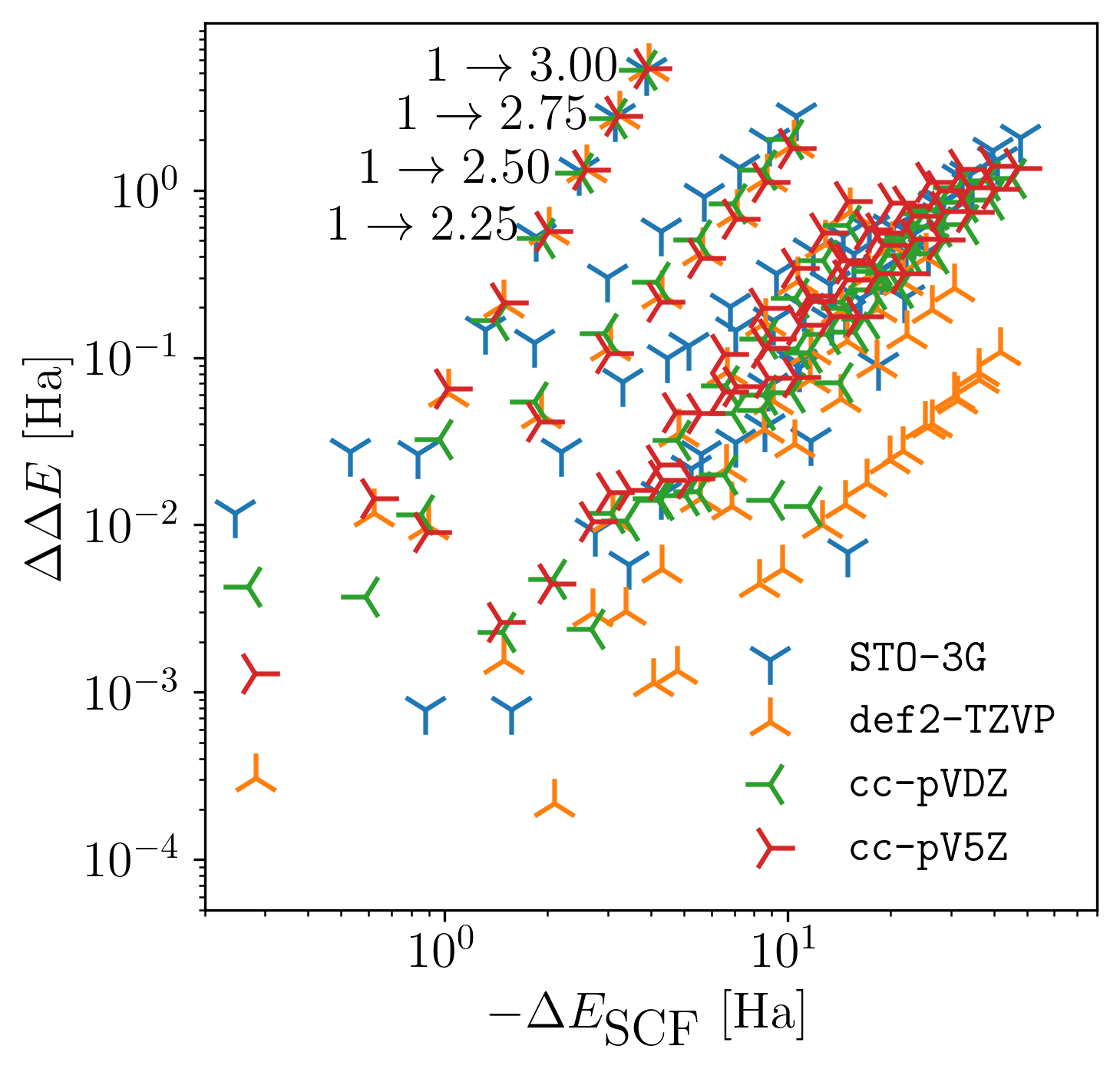}%
        \label{fig:monoatomics_proof_KrOrXe}%
    }
    \caption{The multi-electron atom: absolute error $\Delta \Delta E = |\Delta E_{\text{SCF}} - \Delta E_{\text{AIT}}|$ between unrestricted Hartree-Fock self-consistent field~(SCF) energy difference $\Delta E_{\text{SCF}}$ and the AIT one, $\Delta E_{\text{AIT}}$, up to and including fifth perturbation order $p$, for $Z_A \in \lbrace 1,2,\dots,8\rbrace$ and $Z_B = Z_A + dZ$ with $dZ \in \lbrace 0.25, 0.50, \dots, 2.00 \rbrace$. The basis functions for an initial atom $Z_A$ are in \ref{fig:monoatomics_proof_ind}: each respective element $Z_A$, and in \ref{fig:monoatomics_proof_KrOrXe}: one of the higher available noble gases, Xe in \texttt{def2-TZVP}, Kr in \texttt{STO-3G}, \texttt{cc-pVDZ}, \texttt{cc-pV5Z}. Using the basis functions of noble gases gains much more accuracy than increasing basis set size. Both figures show cases of diverging $\Delta E_{\text{AIT}}$ ($Z_A = 1 \rightarrow Z_B \in \lbrace 2.25,2.50,2.75,3.00 \rbrace$). Note that the energy differences from SCF-calculations in the hydrogen-like atom diverge for $Z_B > 2$, while the ones from analytical computation converge~(Fig.~\ref{fig:hydrogen_proof}). Among the basis sets tested for AIT were the families of Pople, Dunning, Dunning Douglas-Kroll, Dunning JK-fitting, Ahlrichs, Lehtola, ANO and STO, all taken from Ref.~\onlinecite{bse}.}
    \label{fig:monoatomics_proof}
\end{figure}

\clearpage

\section{Performance of noble gases as basis functions}
\label{sec:noble_gas_performances}
\begin{figure}[h!]
    \centering
    \includegraphics[width=0.7\textwidth]{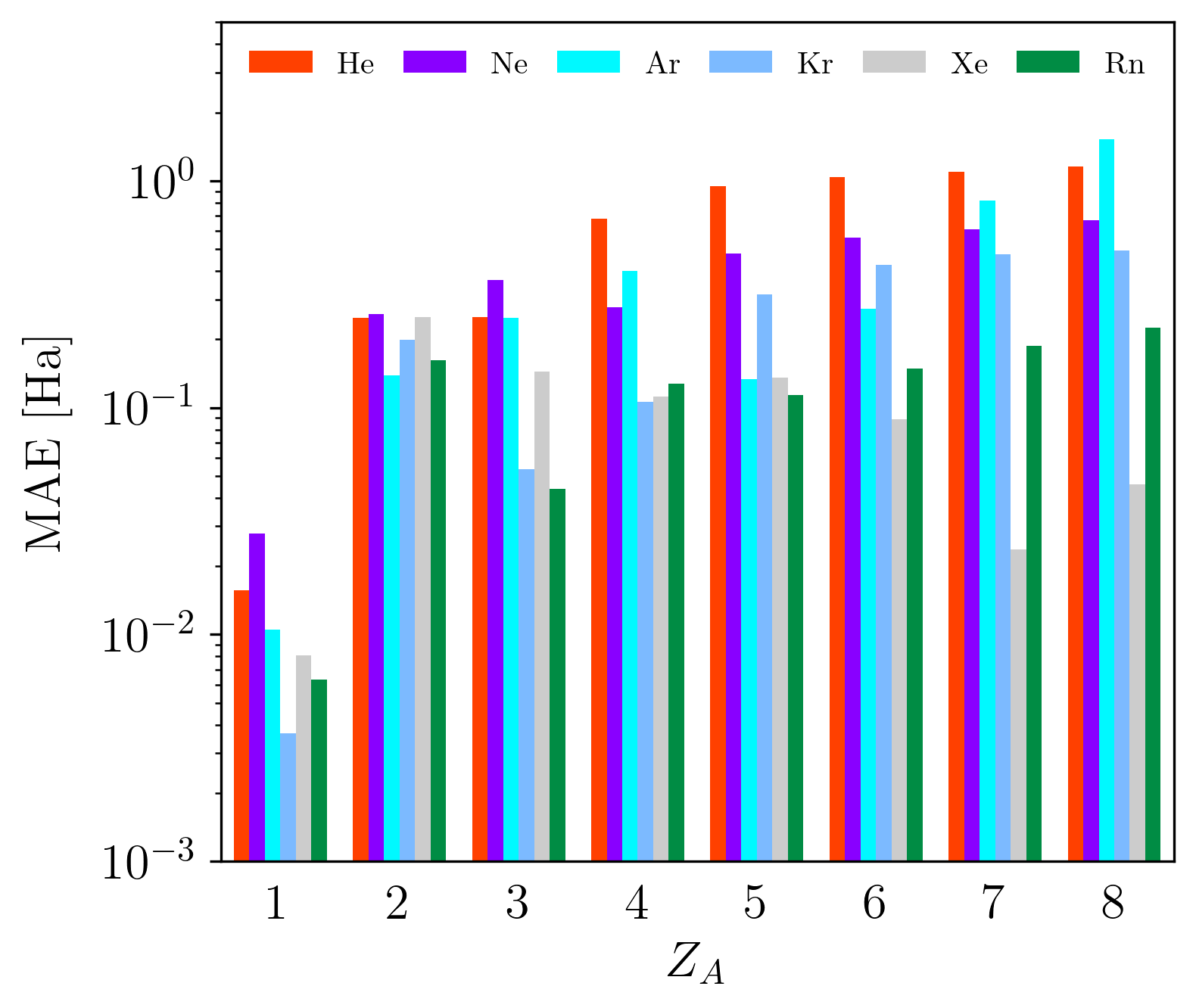}
    \caption{The multi-electron atom: mean average error (MAE) of the same transmutations as in Fig.~\ref{fig:monoatomics_proof} ($Z_A \in \lbrace 1,2,\dots,8\rbrace$ and $Z_B = Z_A + dZ$ with $dZ \in \lbrace 0.25, 0.50, \dots, 2.00 \rbrace$.), excluding divergent ones ($1 \rightarrow Z_B \in \lbrace 2.25, 2.50,2.75,3.00 \rbrace$). First, Kr and Rn appear as more viable candidates but especially for higher $Z_A$, Xe prevails.}
    \label{fig:noble_gas_performances}
\end{figure}

The choice to use the basis functions of Xe in the basis set \texttt{def2-TZVP} is established by the mean average error (MAE) of the six available noble gases (Fig.~\ref{fig:noble_gas_performances}).

\clearpage
\section{AIT's performance in multi-electron atoms for non-integer $\mathbf{Z_A, Z_B}$}

\begin{figure}[b]
    \centering
    \includegraphics[width=0.5\textwidth]{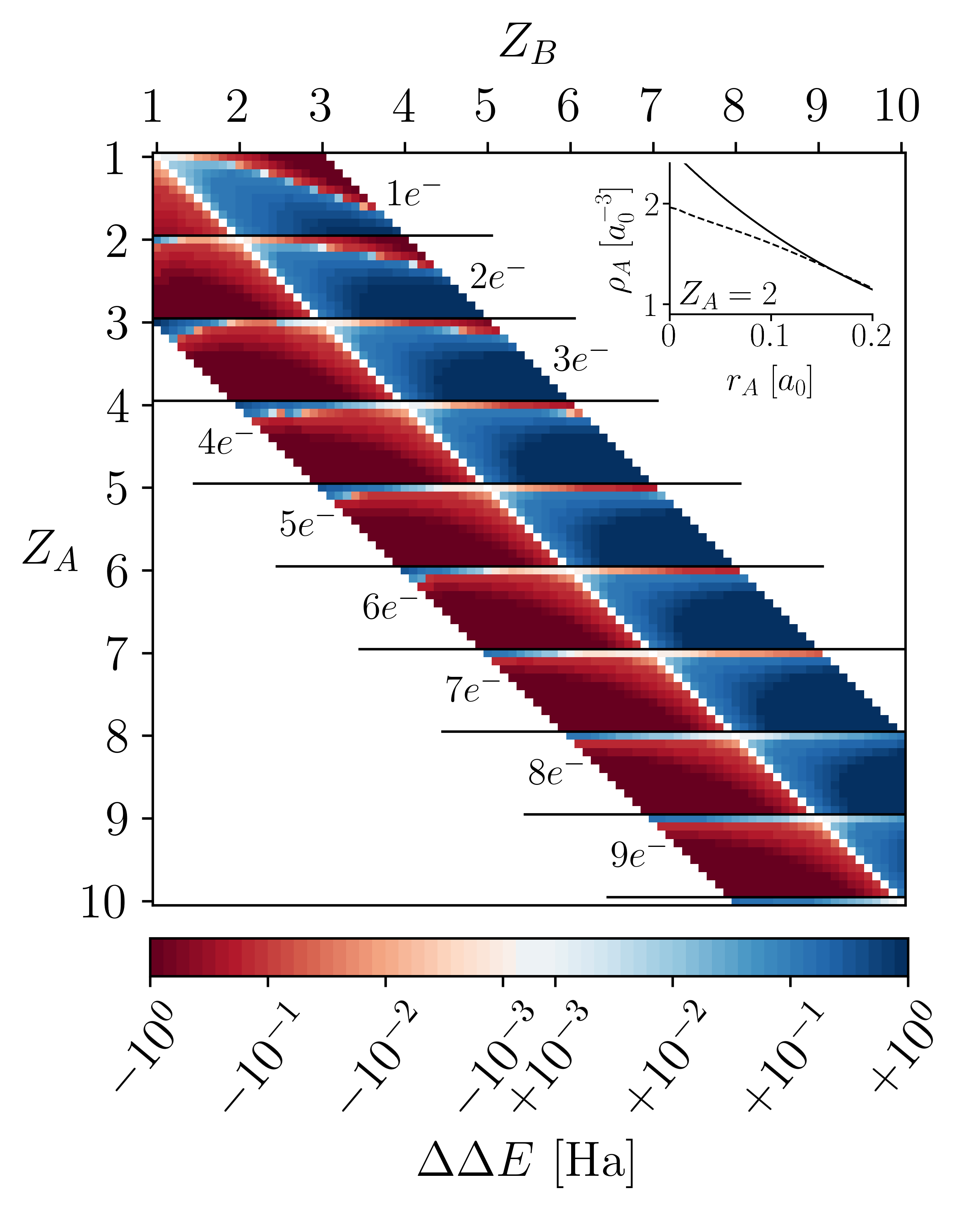}
    \caption{AIT based predictions for multi-electron atoms: Error $\Delta \Delta E = \Delta E_{\text{SCF}} - \Delta E_{\text{AIT}}$ between unrestricted Hartree-Fock SCF energy difference $\Delta E_{\text{SCF}}$ and AIT estimate $\Delta E_{\text{AIT}}$ up to and including fifth perturbation order $p$. Initial electron densities were obtained for $Z_A \in \lbrace 1,1.1,\dots,10.0\rbrace$, and final nuclear charges considered include $Z_B = Z_A \pm dZ$ with $dZ \in \lbrace 0.1, 0.2, \dots, 2.0 \rbrace$.}
    \label{fig:deviation_monoatomic_nonints}
\end{figure}

As a generalization of Fig.~\ref{fig:deviation_monoatomic}, we quantify the absolute error $\Delta \Delta E = \Delta E_{\text{SCF}} - \Delta E_{\text{AIT}}$ for iso-electronic interpolations where $1 \leq Z_A, Z_B \leq 10$, this time without limiting ourselves to integer nuclear charges. Again, we have used the best-performing basis set, \texttt{def2-TZVP} with the basis functions of Xe.
Fig. \ref{fig:deviation_monoatomic_nonints} displays prediction errors as a heat map for all atoms $Z_A \in \lbrace 1.0, 1.1, \dots, 10.0 \rbrace$, with
$Z_B = Z_A \pm dZ$ with $dZ \in \lbrace 0.1, 0.2, \dots 2.0 \rbrace$.
The number of electrons $N_e$ of the initial atom $Z_A$ increases in steps of $1e^-$ such that the initial atom's overall charge $Q_A$ never exceeds $+1$ ($0 \leq Q_A = Z_A - N_e < 1$). 
Note the stripes of sudden increase in accuracy at integer $Z_A$ where $Q_A = 0$, whereas in between integers ($0 < Q_A < 1$), accuracy increasingly worsens, for charged initial atoms, until the next integer~$Z_A$ is reached, another electron is added, and charge-neutrality is recovered ($Q_A = 0$).
Here, the sensitivity to errors in the electron density becomes even more apparent than in Fig.~\ref{fig:deviation_monoatomic}, as SCF solutions between non-integer $Z_A, Z_B$ suffer the most in accuracy.
Furthermore, consider the stripes' curvature (best seen at small $Z_A$) 
where the errors from underestimation of densities coincide with the incipient divergence from AIT due to large $\Delta Z/Z_A$ resulting
in a cancellation of errors. 

\clearpage
\section{Software}
The calculations of any self-consistent field energy or electron density are performed with the restricted/unrestricted Hartree-Fock method of \texttt{PySCF}~\cite{PySCF1, PySCF2} for atoms with even/odd electron numbers. Further software for the purpose of data generation (i.e. math libraries, integration algorithms, numerical tools) in this letter are provided by the \texttt{Python}-packages \texttt{NumPy}~\cite{numpy}, \texttt{SciPy}~\cite{scipy}, \texttt{Numba}~\cite{numba}. Basis sets were provided via \texttt{basis\_set\_exchange}~\cite{bse}. Visualizations have been created using \texttt{Matplotlib}~\cite{matplotlib}.
\end{document}